%% file: DoubleGaussPRD.tex
\documentclass[12pt,aps,prd,preprint,groupedaddress,superscriptaddress,amsmath,amssymb]{revtex4-1}
\usepackage{natbib}
\usepackage{amsmath}    
\usepackage{subfig} 
\usepackage{amssymb}
\usepackage{epsfig}
\usepackage{ifthen}
\usepackage{rotating}
\usepackage{ctable}
\usepackage{graphicx}
\usepackage{dcolumn}
\usepackage{bm}
\usepackage{footnote}
\usepackage{multirow}

\usepackage{color}
\usepackage{eso-pic} 

\usepackage{textcomp}

\usepackage[utf8]{inputenc}
\usepackage{mathrsfs}

 \usepackage{graphicx}                                  
\DeclareGraphicsExtensions{.eps,.ps,.eps.gz,.ps.gz}    
 
\usepackage[linktocpage,hidelinks]{hyperref}
\usepackage{breakurl}
\usepackage{url}

\makeatletter

\renewcommand\l@subsection{\@dottedtocline{2}{1.em}{3.5em}}
\renewcommand*\l@figure{\@dottedtocline{1}{2em}{4.em}}
\let\l@table\l@figure
\makeatother

\include{afbcommands}

\begin{document}
\thispagestyle{empty}


\title{On the Forward-Backward Asymmetry of Leptonic Decays of $\bf{\ttbar}$ at the Fermilab Tevatron}
\author{Ziqing Hong}
\email{zqhong@fnal.gov}
\affiliation{Mitchell Institute for Fundamental Physics and Astronomy, Texas A\&M University}
\author{Ryan Edgar}
\affiliation{University of Michigan}
\author{Sarah Henry}
\affiliation{Mitchell Institute for Fundamental Physics and Astronomy, Texas A\&M University}
\author{David Toback}
\affiliation{Mitchell Institute for Fundamental Physics and Astronomy, Texas A\&M University}
\author{Jonathan S. Wilson}
\affiliation{University of Michigan}
\author{Dante Amidei}
\affiliation{University of Michigan}

\begin{abstract}
We report on a study of the measurement techniques used to determine the leptonic forward-backward asymmetry of top anti-top quark pairs in Tevatron experiments with a proton anti-proton initial state. Recently it was shown that a fit of the differential asymmetry as a function of $\qeta$ (where $q_{l}$ is the charge of the lepton from the cascade decay of the top quarks and $\eta_{l}$ is the final pseudorapidity of the lepton in the detector frame) to a hyperbolic tangent function can be used to extrapolate to the full leptonic asymmetry. We find this empirical method to well reproduce the results from current experiments, and present arguments as to why this is the case. We also introduce two more models, based on Gaussian functions, that better model the $\qeta$ distribution. With our better understanding, we find that the asymmetry is mainly determined by the shift of the mean of the $\qeta$ distribution, the main contribution to the inclusive asymmetry comes from the region around $|\qeta| = 1$, and the extrapolation from the detector-covered region to the inclusive asymmetry is stable via a multiplicative scale factor, giving us confidence in the previously reported experimental results.
\end{abstract}

\date{\today}

\maketitle


%
%
%


\section{\label{sec:Intro}Introduction}
Recent measurements of the forward-backward asymmetry ($\afb$) of top anti-top quark pair ($\ttbar$) production in proton anti-proton collisions with $\sqrt{s} = 1.96$~TeV at the Fermilab Tevatron~\cite{Aaltonen:2012it,d0_afb_prd2011,CDF:2013gna} have shown anomalously large values compared to the predictions from the standard model (SM) of particle physics at next-to-leading order (NLO)~\cite{Bernreuther:2012sx,*PhysRevLett.81.49}. This is of great interest as new particles or interactions could cause the $\afb$ of $\ttbar$ ($\afbtt$) to be different from SM-only predictions~\cite{Jung:2010yn,*Jung2010238,*Frampton2010294,*jhep05(2011)070,*Chen:2010hm,*PhysRevD.82.094011,*PhysRevD.82.071702,
*PhysRevD.82.094009,*PhysRevD.82.034026,*PhysRevD.81.114004,*PhysRevD.81.055009,*PhysRevD.81.015004,
*PhysRevD.81.034012,*PhysRevD.82.034034,*PhysRevD.81.014016,*PhysRevD.81.113009,*PhysRevD.78.094018,
*PhysRevD.80.051701,*jhep11(2010)039,*Cheung2009287}. An alternative observable that could be also affected is the forward-backward asymmetry of the leptons from the cascade decay of the top quarks, the so-called leptonic forward-backward asymmetry ($\afblep$)~\cite{Bernreuther201090}. In addition, $\afblep$ can deviate further from its SM prediction in the scenarios that the top quarks are produced with a certain polarization. For example, resonant production of $\ttbar$ via a hypothesized axigluon could cause the $\afbtt$ to vary from its SM value; while different chiral couplings between the axigluons and the top quarks could produce the same value of $\afbtt$, but very different values of $\afblep$~\cite{PhysRevD.87.034039}.

For a sample of $\ttbar$ events that decay into one or more charged leptons, the $\afblep$ is defined as
\begin{align}
	\afblep &= \frac{N(\qeta > 0) - N(\qeta < 0)}{N(\qeta > 0) + N(\qeta < 0)} 
	\label{eqn:afblep}
\end{align} 
where $N$ is the number of charged leptons (electrons or muons) in the sample, $q_{l}$ is the lepton charge, and $\eta_{l}$ is the pseudorapidity of the charged lepton. The measurement of $\afblep$ has been done in both the lepton+jets final state (where only one $W$ boson from the top quarks decays leptonically) and the dilepton final state (where both $W$ bosons decay leptonically) at both the CDF~\cite{Aaltonen:2013vaf, CDF11035} and the D0~\cite{Abazov:2013wxa,D0LJafb} experiments using different methods. Of critical importance for this measurement is the methodology to extrapolate from the finite coverage of the experiments ($|\eta_{l}|<1.25$ and $|\eta_{l}|<2.0$ in the lepton+jets and dilepton final states respectively) to the full pseudorapidity range (inclusive) parton-level result. A method, first proposed in Ref.~\cite{Aaltonen:2013vaf}, is to decompose the measured $\qeta$ distribution into a symmetric part ($\Sqeta$ term) and an asymmetric part (differential forward-backward asymmetry, $\Aqeta$ term). Studies indicated that the $\Sqeta$ term was nearly model independent; using a distribution estimated with any sample of simulated events only introduces a small systematic uncertainty. Equally important is that the $\Aqeta$ term was found to vary significantly from model to model as a function of $\afblep$, allowing for a measurement; this part is measured directly from data. Interestingly, empirical studies showed that a hyperbolic tangent function could be used to model the $\Aqeta$ term with a measurement bias that was negligible compared to the other uncertainties.

In this article, we first briefly describe the parametrization introduced by Ref.~\cite{Aaltonen:2013vaf}, then introduce more detailed studies of the parton level $\qeta$ distribution to both understand why the hyperbolic tangent function works so well, and to see what improvements could be made with a better understanding. We find that the $\qeta$ distribution is actually well described by a double-Gaussian distribution, where the asymmetry arises from a shift in the mean of the distribution. We conclude this manuscript with the implications of this modeling, as well as our thoughts for future measurements.

\section{\label{sec:tanhParametrization}Leptonic \texorpdfstring{$\mathbf{\afb}$}{} Measurements at the Tevatron}

\begin{figure}[hbtp]
  \begin{center}
  	\subfloat[]{\includegraphics[width=\columnwidth]{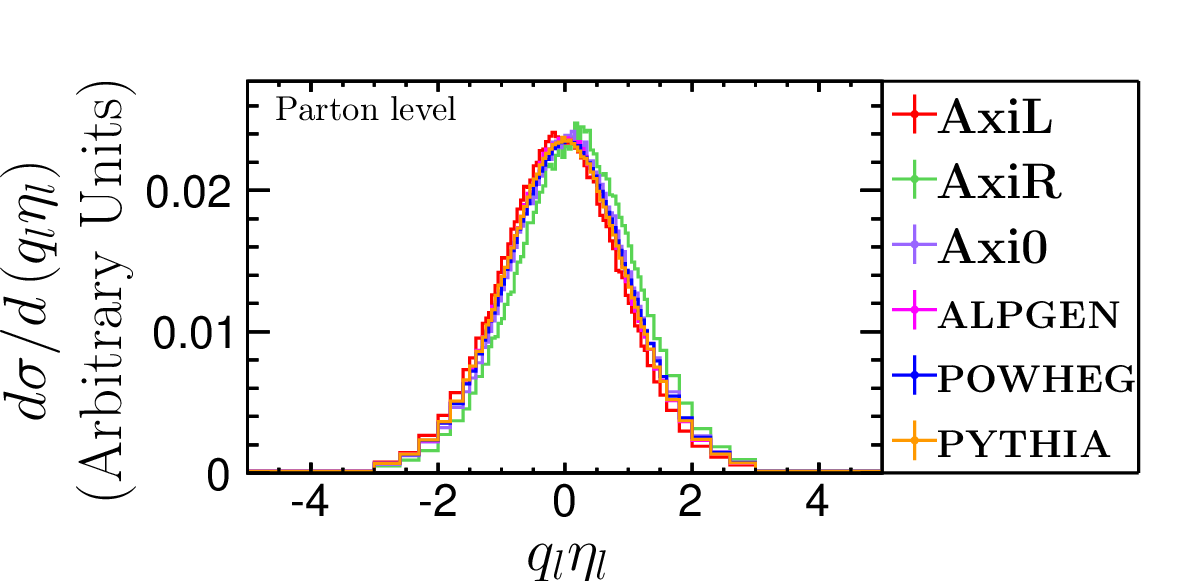}  }\\
  	\subfloat[]{\includegraphics[width=\columnwidth]{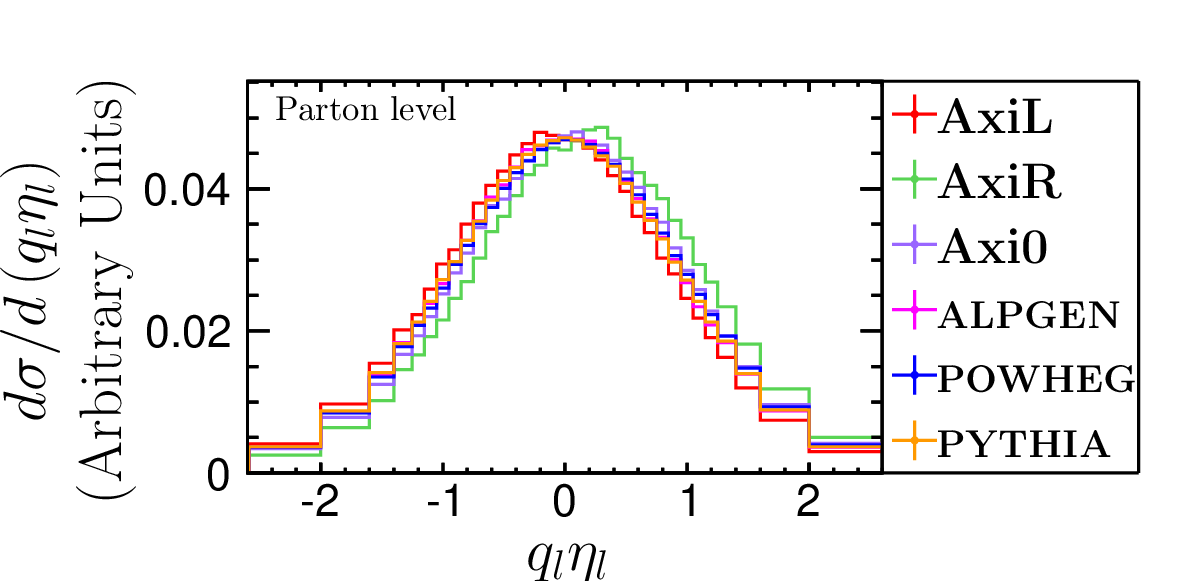}  }  	
  \end{center}
  	\caption{The $\qeta$ distribution of charged leptons produced from $\ttbar$ cascade decay from simulations with various physics models at parton level, before any selection requirements. In (b) only the range between -2.5 and 2.5 is shown.}
  	\label{fig:qeta_hepg}
  \end{figure}

To study the $\qeta$ distribution with different physical scenarios, we used six benchmark Monte Carlo (MC) simulated samples. To model the SM we consider two leading-order (LO) SM samples generated by \textsc{pythia}~\cite{Sjostrand:2006za} and \textsc{alpgen}~\cite{Mangano:2002ea}, and for NLO effects we use a sample generated with \textsc{powheg}~\cite{Frixione:2007nw,Nason:2004rx,Frixione:2007vw,Alioli:2010xd}; we note that the \textsc{powheg} sample does have quantum chromodynamics (QCD) effects, but does not have electroweak (EWK) effects~\cite{Antunano:2007da,*PhysRevD.84.093003,*Manohar2012313,*jhep012012063}. To test the measurement on a larger range of $\afblep$, we consider three samples with physics beyond the SM, with a class of relatively light and wide axigluons (m = 200~$\gevcc$, $\Gamma = 50~\gevcc$) with left-handed, right-handed and axial flavor-universal couplings to the quarks~\cite{PhysRevD.87.034039}, generated with \textsc{madgraph}~\cite{Alwall:2007st}. These are chosen as they all predict an $\afbtt$ value that is close to the value observed at CDF~\cite{Aaltonen:2012it}, but give very different predictions of $\afblep$~\cite{PhysRevD.87.034039,CDF11035}. The $\qeta$ distributions at parton level for all six benchmark $\ttbar$ samples are shown in Fig.~\ref{fig:qeta_hepg}. The $\afblep$ values predicted by the samples span the range of $-0.1<\afblep<0.2$ and are listed in Table~\ref{table:SigMC} along with a full NLO SM calculation, together with the results of the measurements from CDF and D0 in both the lepton+jets and dilepton final states. We note that the measurement from D0 in the lepton+jets final state is limited to the region where $|\qeta| < 1.5$. Later in this article we provide a stable extrapolation of the $\afblep$ in this region to the inclusive $\afblep$.

\begin{table*}[hbtp]
\caption{A collection of different predictions and measurements of $\afblep$ from various sources. The uncertainties for the simulated samples are statistical only. The uncertainty for the NLO SM calculation is due to the variation in the scales in the calculation. The uncertainties for the CDF and D0 results are the overall uncertainties from the measurements.}
\begin{tabular}{cccc}
\hline 
Source & $\afblep$  & Description \\ 
\hline \hline
{\textsc{madgraph}}   &\multirow{2}{*}{-0.063$\pm$0.002} & Tree-level left-handed axigluon\\
(AxiL)&  &(m = 200~$\gevcc$, $\Gamma = 50~\gevcc$) \\ 
\hline
{\textsc{madgraph}}   & \multirow{2}{*}{0.151$\pm$0.002} & Tree-level right-handed axigluon\\
(AxiR)&  & (m = 200~$\gevcc$, $\Gamma = 50~\gevcc$) \\ 
\hline
{\textsc{madgraph}}   & \multirow{2}{*}{0.050$\pm$0.002} & Tree-level unpolarized axigluon\\
(Axi0)&  & (m = 200~$\gevcc$, $\Gamma = 50~\gevcc$) \\ 
\hline
\textsc{alpgen} & 0.003$\pm$0.001 & Tree-level Standard Model \\ 
\hline
\textsc{pythia} & 0.000$\pm$0.001 & LO Standard Model \\ 
\hline
\textsc{powheg} & 0.024$\pm$0.001 & NLO Standard Model with QCD corrections \\ 
\hline
\hline
Calculation & 0.038$\pm$0.003 & NLO SM with QCD and EWK corrections~\cite{Bernreuther:2012sx,*PhysRevLett.81.49} \\
\hline
\hline
\multirow{3}{*}{CDF} & $0.094^{+0.032}_{-0.029}$ & Lepton+jets\\
                     & $0.072 \pm 0.060$         & Dilepton\\
                     & $0.090^{+0.028}_{-0.026}$ & Combination\\
\hline
\multirow{2}{*}{D0}  & $0.047^{+0.025}_{-0.027}$ & Lepton+jets, $|\qeta|<1.5$\\
                     & $0.044 \pm 0.039$         & Dilepton\\
\hline\hline
\end{tabular}
\label{table:SigMC}
\end{table*}

As described in Ref.~\cite{Aaltonen:2013vaf,CDF11035}, the $\qeta$ distribution of the leptons can be decomposed into an $\Sqeta$ term and an $\Aqeta$ term using the following formulas in the range $\qeta \geq 0$:
\begin{subequations}
\begin{align}
	\Sqeta &= \frac{\mathcal{N}(\qeta)+\mathcal{N}(-\qeta)}{2} \label{eqn:S},~\rm{and}\\
	\Aqeta &= \frac{\mathcal{N}(\qeta)-\mathcal{N}(-\qeta)}{\mathcal{N}(\qeta)+\mathcal{N}(-\qeta)},\label{eqn:A}
\end{align}
\label{eqn:decomposition}
\end{subequations}
\noindent where $\mathcal{N}(\qeta)$ represents the number of events as a function of $\qeta$. With this, the $\afblep$ defined in Eq.~(\ref{eqn:afblep}) can be rewritten in terms of $\Sqeta$ and $\Aqeta$ as:
\begin{equation}
\begin{aligned}
	\afblep &= \frac{\int_{0}^{\infty}\mathrm{d}x~ [\mathcal{A}(x) \cdot \mathcal{S}(x)] }
	{\int_{0}^{\infty} \mathrm{d}x' ~\mathcal{S}(x')} .\label{eqn:S_A_AFB}
\end{aligned}
\end{equation}

\begin{figure}[htp]
\begin{center}
	\subfloat[\label{fig:s_qeta_hepg}]{\includegraphics[width=\columnwidth]{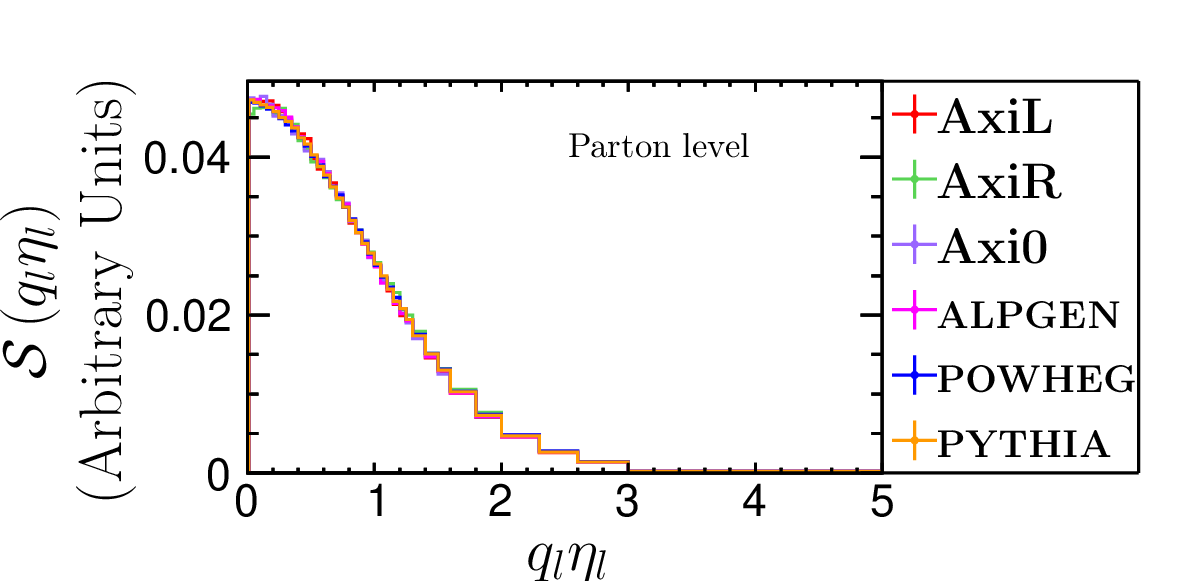}}\\
	\subfloat[\label{fig:a_qeta_hepg}]{\includegraphics[width=\columnwidth]{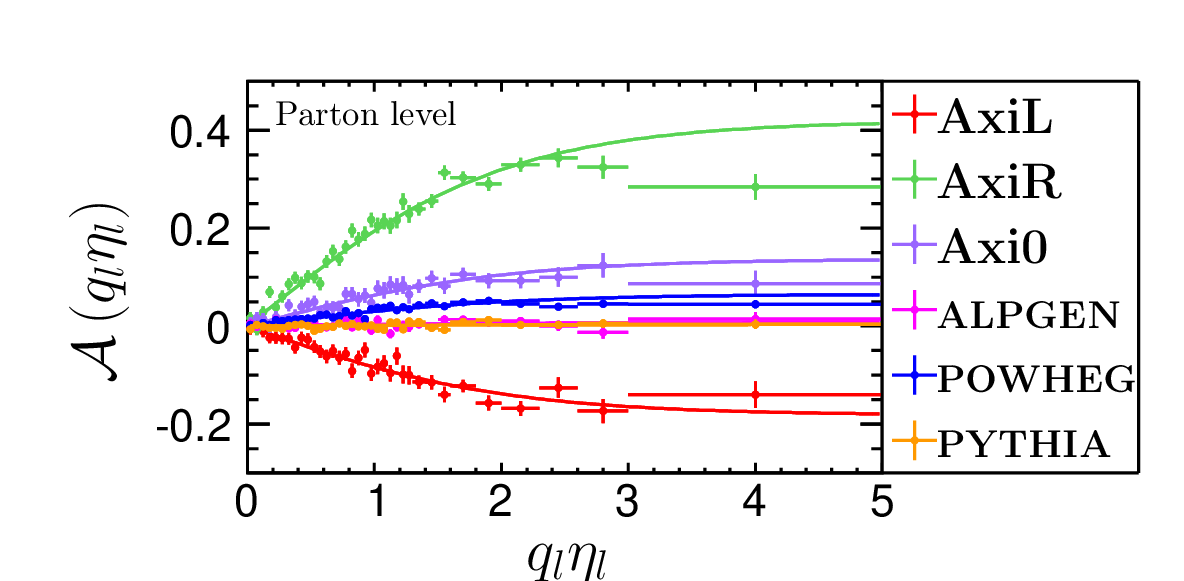}}
	\caption{The $\Sqeta$ term (a) and the $\Aqeta$ term (b) of the $\qeta$ distribution from various physics models. The lines in (b) correspond to the best fits from the $\atanh$ model.}
	\label{fig:s_qeta_a_qeta_hepg}
\end{center}
\end{figure}

\begin{figure}[htp]
\begin{center}
	\includegraphics[width=\columnwidth]{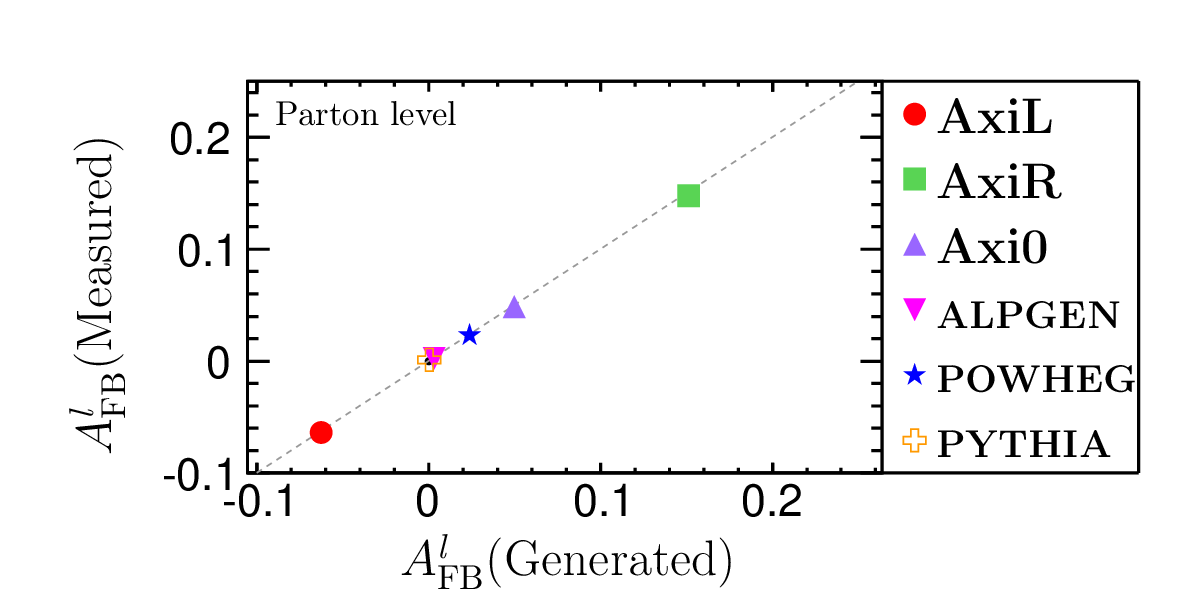}
	\caption{A comparison between the predicted $\afblep$ from simulations and the $\afblep$ as measured using the $\atanh$ parametrization with parton level information from $|\qeta|<2.0$. The dashed line indicates the location of the equal values, while the points are superimposed at their measured locations. All the points lie along the line within uncertainties.}
	\label{fig:afb_diag_hepg}
\end{center}
\end{figure}

The $\Sqeta$ term and the $\Aqeta$ term distributions from the benchmark samples are shown in Fig.~\ref{fig:s_qeta_hepg} and Fig.~\ref{fig:a_qeta_hepg}, respectively. We can readily see that the variation of the $\Sqeta$ term among the benchmark $\ttbar$ samples is small, so choosing any one of them for the measurement introduces an uncertainty that is tiny compared to the dominant uncertainties. We will come back to the small differences for $\qeta < 0.2$ and show why they do not have much effect on the measurement. On the other hand, the $\Aqeta$ term varies significantly from model to model. The $\Aqeta$ term has been well described in the region $|\qeta|<2.0$ using the ansatz of
\begin{align}
	\Aqeta &= a \cdot \mathrm{tanh}\left(\frac{1}{2} \qeta\right)\label{eqn:tanh}
\end{align}
where $a$ is a free parameter that is directly related to the final asymmetry. Best fits of the data to the $\atanh$ model from Eq.~(\ref{eqn:tanh}) are also shown in Fig.~\ref{fig:a_qeta_hepg}. While the $\Aqeta$ term is well modeled in the region where $\qeta < 2.5$, it is not as good above 2.5. The comparison between the predicted $\afblep$ and the $\afblep$ obtained with a measured value of $a$ in Eq.~(\ref{eqn:tanh}) from the $\Aqeta$ term (restricting the fit within the region $\qeta<2.0$ to simulate a detector) is shown in Fig.~\ref{fig:afb_diag_hepg}. The differences are on the order of a fraction of a percent, which is tiny compared to the dominant uncertainties listed in Table~\ref{table:SigMC}~\cite{Aaltonen:2013vaf,CDF11035}. 

While the methodology works well, the parametrization of Eq.~(\ref{eqn:tanh}) is purely empirical. In the following sections, we provide a partial explanation of where the hyperbolic tangent functional form comes from as well as a better parametrization the new understanding leads us to.

\clearpage

We will further see that the choice of the observable $\afblep$ is advantageous because it uses the precise measurement of the value of $\qeta$ for each lepton. Thus, the bin-to-bin migration of events due to detector smearing is small, and has no measurable effect on the final value of $\afblep$. However, we will also see that judicious choices of the binning, especially at large $|\qeta|$, are important when using fitting and extrapolation techniques.


\section{\label{sec:singleGaussian}Single and Double Gaussian Modeling}

\begin{figure}[hbtp]
	\begin{center}
		\subfloat[]{\includegraphics[width=\linewidth]{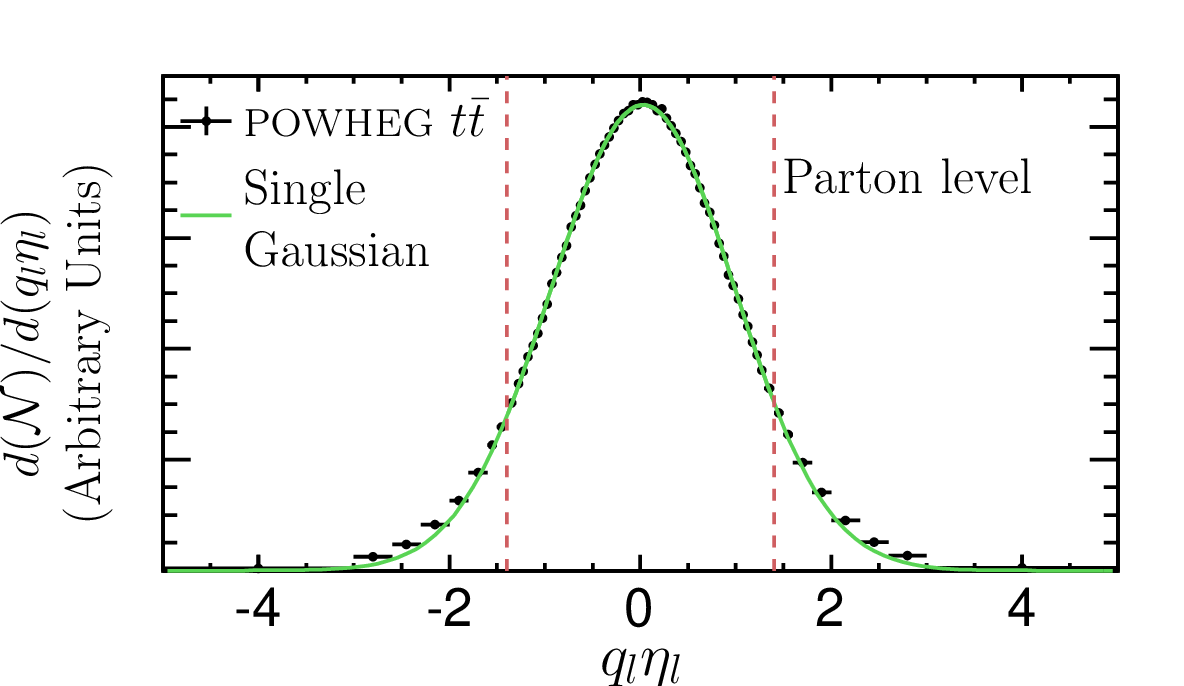}\label{fig:single_gaussian_a}}\\
		\subfloat[]{\includegraphics[width=\linewidth]{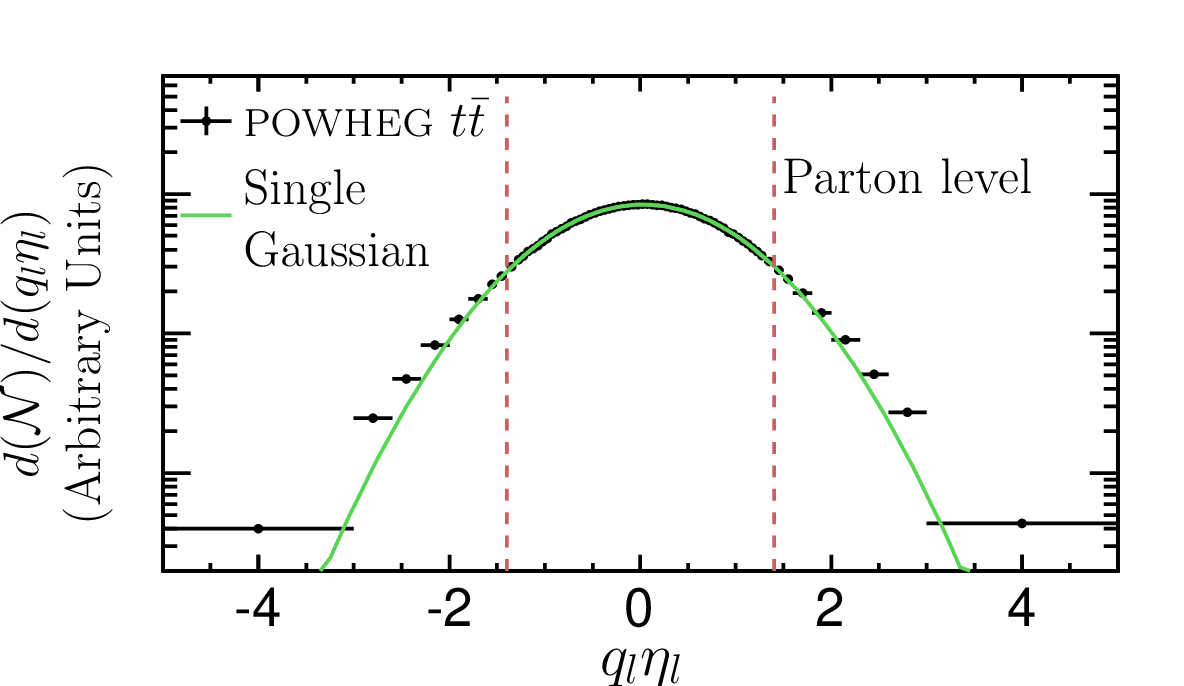}\label{fig:single_gaussian_b}}
		\caption{The $\qeta$ distribution from the \textsc{powheg} $\ttbar$ sample at parton level, with a fit to a single Gaussian function in the region $|\qeta|<1.4$ (indicated by the dashed lines). Note that the agreement is not good for $|\qeta|>1.4$.}
		\label{fig:single_gaussian}
	\end{center}
\end{figure}

The $\qeta$ distributions in Fig.~\ref{fig:qeta_hepg} appear to be roughly Gaussian distributed with a non-zero mean. However, the Gaussian model is only good in the small-$|\qeta|$ region. Figure~\ref{fig:single_gaussian} shows the $\qeta$ distribution at parton level from the \textsc{powheg} $\ttbar$ sample with a fit to a Gaussian function, but with the fit restricted to $|\qeta|<1.4$. Note that the fit is not good for $|\qeta|>1.4$. This simple model is clearly insufficient.

Before moving on to a better model, we use this simple model to illustrate the methodology. We note that the number of events in the interval ($\qeta$, $\qeta$+$\mathrm{\delta}(\qeta)$) can be readily calculated using
\begin{widetext}
\begin{equation}
	\begin{aligned}
	\mathcal{N}(\qeta,\qeta+\mathrm{\delta}(\qeta))&=\int_{\qeta}^{\qeta+\mathrm{\delta}(\qeta)} \mathrm{d}x~C \cdot \mathrm{Exp}(-\frac{(x-\mu)^{2}}{2\sigma^{2}})\\
	&=C \cdot \mathrm{Exp}(-\frac{(\qeta-\mu)^{2}}{2\sigma^{2}})~\mathrm{\delta}(\qeta)~\text{,when}~\mathrm{\delta}(\qeta)\rightarrow 0,
	\end{aligned}
\end{equation}
\end{widetext}
where $C$ is a normalization constant, $\mu$ is the mean of the distribution and $\sigma$ is the width of the distribution. We can then calculate $\Aqeta$ with this function:
\begin{widetext}
\begin{equation}
	\begin{aligned}
  	\Aqeta &= \lim_{\mathrm{\delta}(\qeta) \rightarrow 0} \frac{\mathcal{N}(\qeta,\qeta+\mathrm{\delta}(\qeta)) - \mathcal{N}(-\qeta-\mathrm{\delta}(\qeta),-\qeta)}{\mathcal{N}(\qeta,\qeta+\mathrm{\delta}(\qeta)) + \mathcal{N}(-\qeta-\mathrm{\delta}(\qeta),-\qeta)}\\
  			&= \frac{\mathrm{Exp}(-\frac{(\qeta-\mu)^{2}}{2\sigma^{2}})-\mathrm{Exp}(-\frac{(-\qeta-\mu)^{2}}{2\sigma^{2}})}{\mathrm{Exp}(-\frac{(\qeta-\mu)^{2}}{2\sigma^{2}})+\mathrm{Exp}(-\frac{(-\qeta-\mu)^{2}}{2\sigma^{2}})}\\
			&= \mathrm{tanh}(\frac{\mu \cdot \qeta}{\sigma^{2}})\label{eqn:singleGaussian}
	\end{aligned}
\end{equation}
\end{widetext}

We note that it has the form of a hyperbolic tangent function, but with the parameter inside the function argument, not an overall scaling factor as in Eq.~(\ref{eqn:tanh}).

\begin{figure}[hbtp]
\begin{center}
\subfloat[]{\includegraphics[width=\columnwidth]{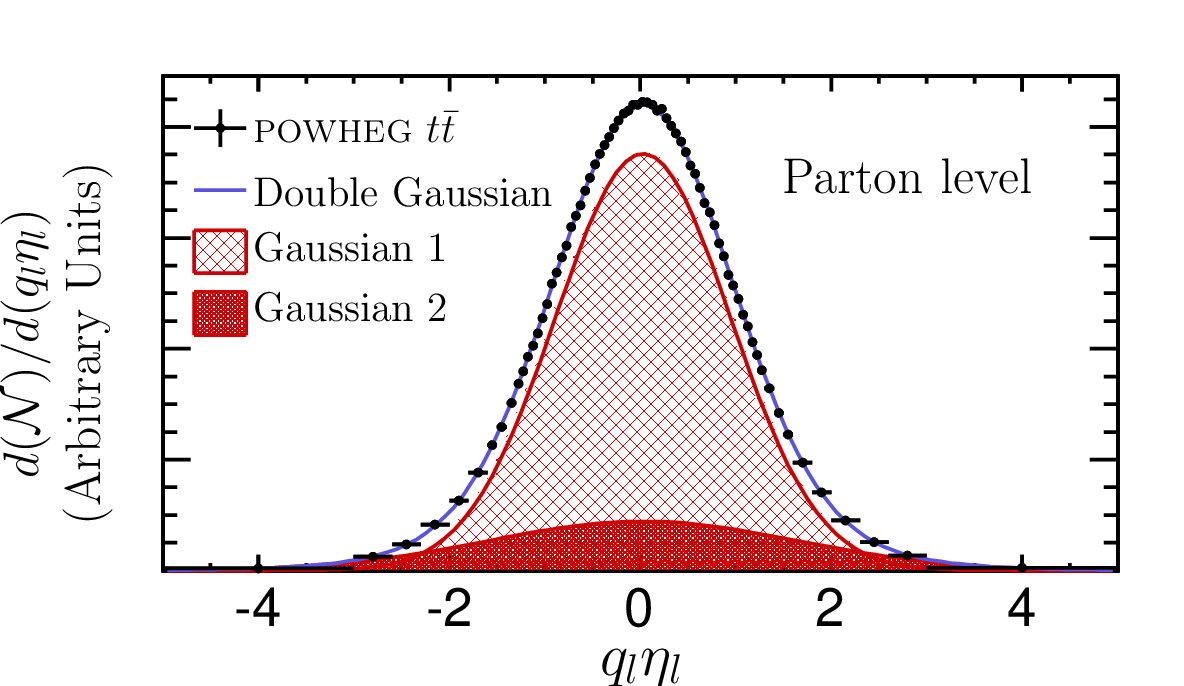}\label{fig:double_gaussian_a}}\\
\subfloat[]{\includegraphics[width=\columnwidth]{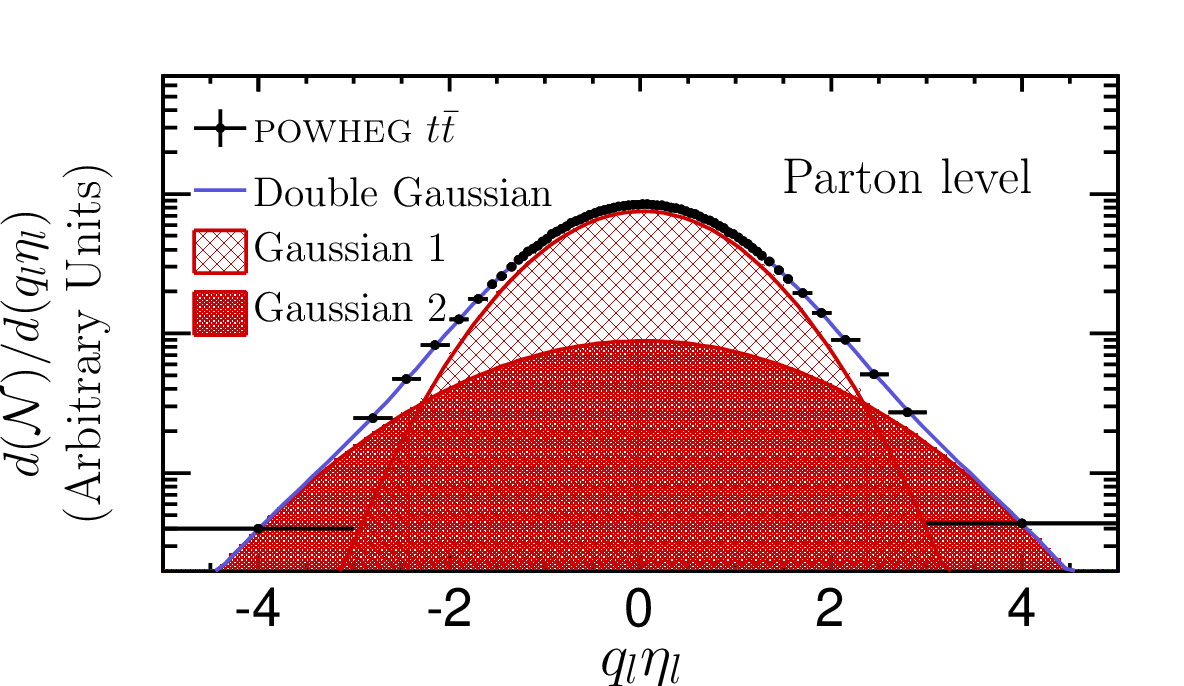}\label{fig:double_gaussian_b}}
\caption{The $\qeta$ distribution from the \textsc{powheg} $\ttbar$ sample at parton level, overlaid with the double-Gaussian fit. Note that both the tails and the central part of the distribution are well described.}
\label{fig:doubleGaussian}
\end{center}
\end{figure}

\begin{figure*}[hbtp]
\begin{center}
	\subfloat[]{\includegraphics[width=0.49\textwidth]{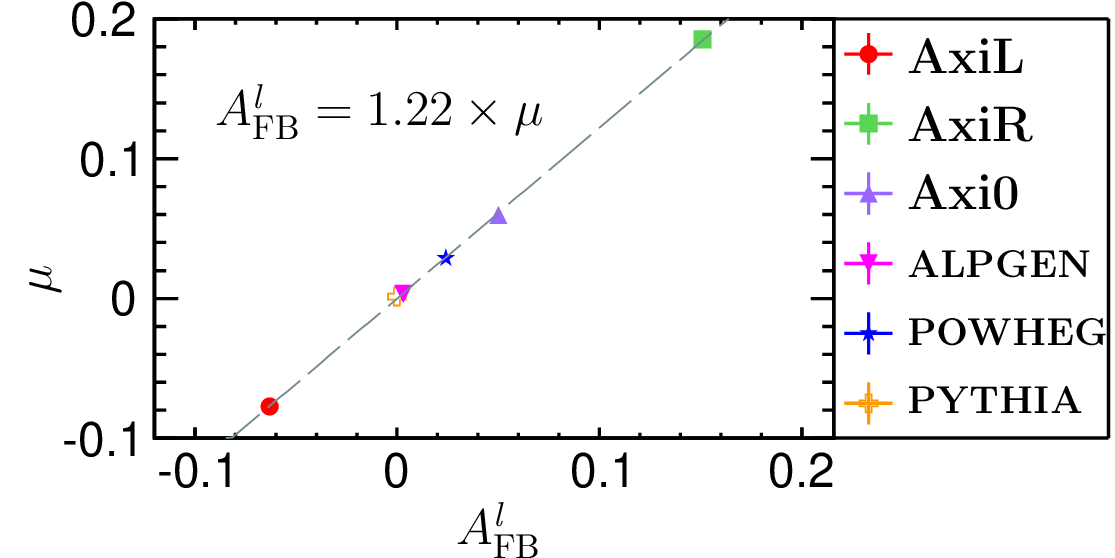}\label{fig:AFB_MU_MC}}
	\subfloat[]{\includegraphics[width=0.49\textwidth]{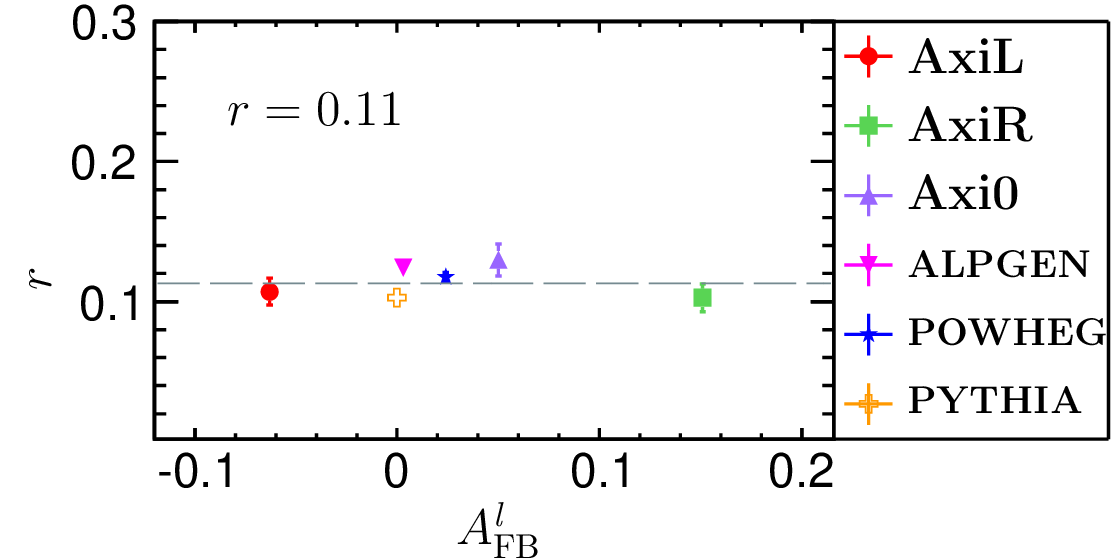}}\\
	\subfloat[]{\includegraphics[width=0.49\textwidth]{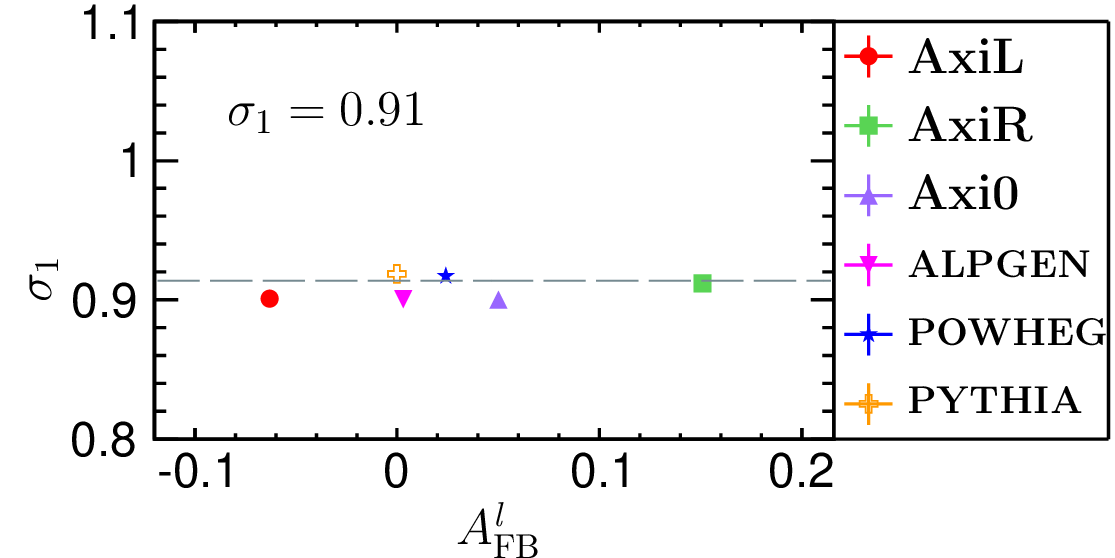}}
	\subfloat[]{\includegraphics[width=0.49\textwidth]{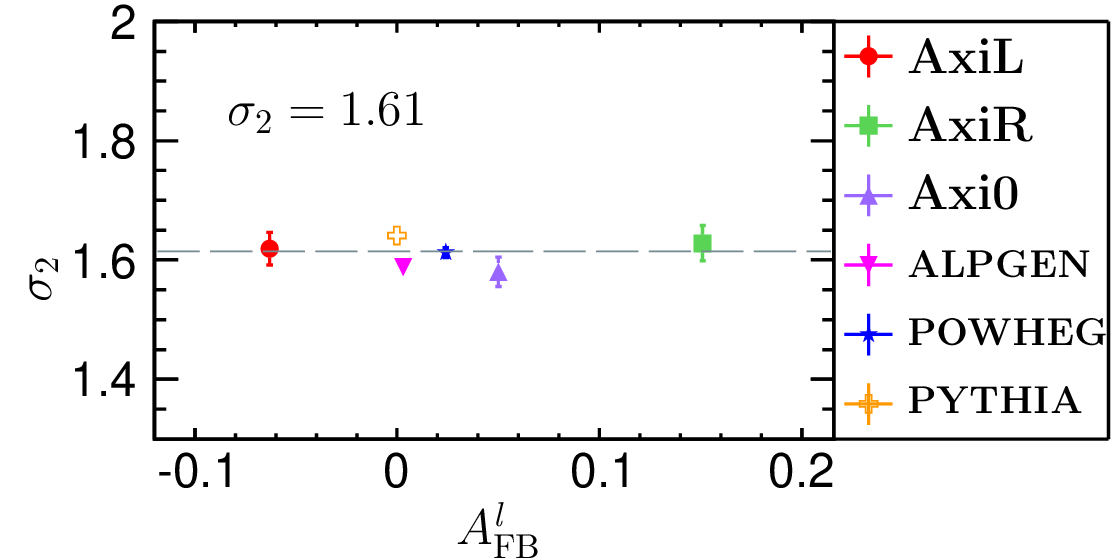}}
	\caption{Fit parameters from our benchmark samples as a function of $\afblep$.}
	\label{fig:doubleGaussianParameters}
\end{center}
\end{figure*}

Since the single Gaussian function works only in the small $|\qeta|$ region, we tried a more sophisticated model, and found that the sum of two Gaussian functions with a common mean works very well at describing the data, even at large values of $\qeta$. We have not uncovered an \textit{a priori} explanation why this should be so, but it appears to be true for all the models we considered~\cite{mangano}. We use the functional form:
\begin{widetext}
\begin{equation}
\frac{\mathrm{d} \mathcal{N}(\qeta)}{\mathrm{d} (\qeta)} = C \cdot \Big( \mathrm{Exp}(-\frac{(\qeta-\mu)^{2}}{2\sigma_{1}^{2}}) + r \cdot \mathrm{Exp}(-\frac{(\qeta-\mu)^{2}}{2\sigma_{2}^{2}}) \Big)\label{eqn:doubleGaussian},
\end{equation}
\end{widetext}
where $C$ is a normalization constant, $r$ is a multiplicative factor that covers the relative normalization of the two components and $\sigma_1$ and $\sigma_2$ are the widths of the two different distributions. Fig.~\ref{fig:doubleGaussian} shows a comparison between the best fit and the parton level data. This functional form works well for all our benchmark signal samples; the two $\sigma$ terms and the $r$ term are very consistent as shown in Fig.~\ref{fig:doubleGaussianParameters}. We find $\sigma_{1} = 0.91$, $\sigma_{2} = 1.61$ and $r = 0.11$. More importantly, the mean ($\mu$) varies significantly from one sample to another, and appears to be linear with $\afblep$. From here on, we assume the two $\sigma$ terms and the $r$ term have the best fit values from the benchmark samples for further studies.

\begin{figure*}[htbp]
\begin{center}
\subfloat[]{\includegraphics[width=0.49\textwidth]{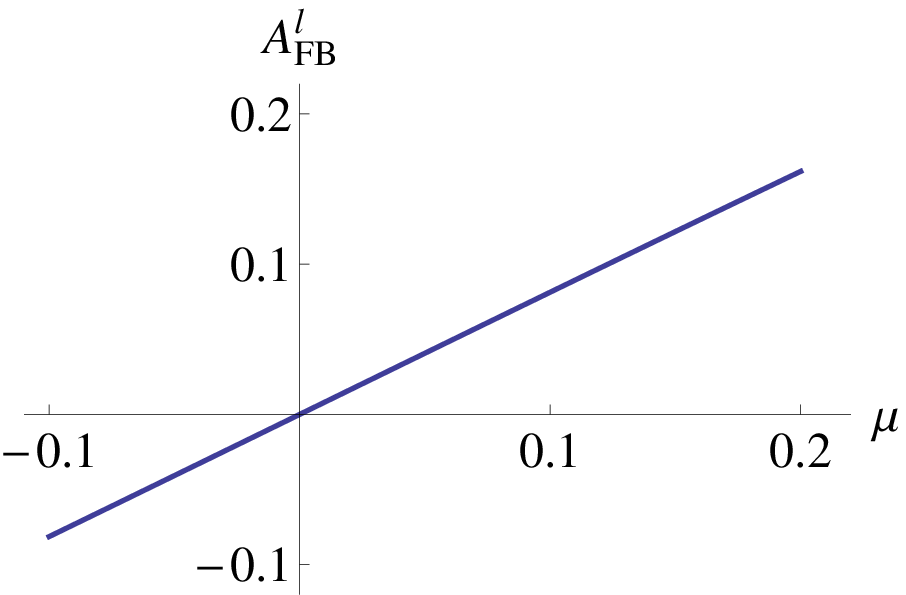}\label{fig:AFB_mu_zoomin}}
\subfloat[]{\includegraphics[width=0.49\textwidth]{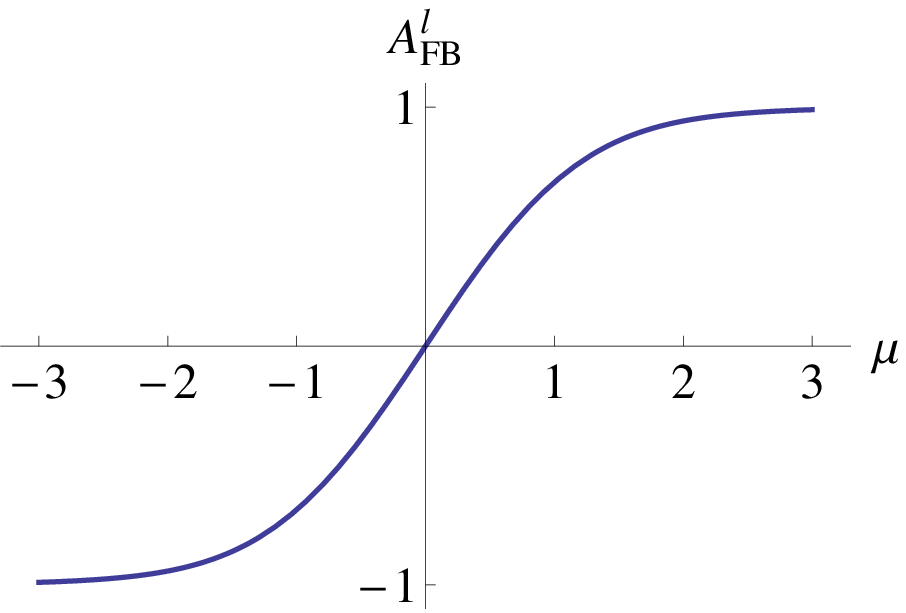}}
\end{center}
\caption{With the double-Gaussian modeling, and constraining the two $\sigma$ values and the $r$ to the best estimated values from the benchmark simulations, $\afblep$ appears to be linear as a function of the mean of the double-Gaussian function in the small $\afblep$ region. In a larger region, $\afblep$ asymptotes to $\pm$1.}
\label{fig:AFB_mu}
\end{figure*}

The double-Gaussian modeling allows for closed form calculations of the $\Sqeta$ and the $\Aqeta$ terms as well as the inclusive $\afblep$ using just the $\mu$, $\sigma_1$, $\sigma_2$ and $r$ parameters. We find the $\Sqeta$ term and the $\Aqeta$ term have the functional forms of
\begin{widetext}
\begin{subequations}
\begin{align}
\Sqeta &=\frac{C}{2} \cdot \left(e^{-\frac{(\qeta-\mu)^2}{2 \sigma_1^2}}+e^{-\frac{(\qeta+\mu )^2}{2 \sigma_1^2}} + r \cdot e^{-\frac{(\qeta-\mu)^2}{2 \sigma_2^2}}+r\cdot e^{-\frac{(\qeta+\mu )^2}{2 \sigma_2^2}}\right)\label{doubleGaussian_S}, \rm{and} \\
\Aqeta &=\frac{e^{-\frac{(\qeta-\mu )^2}{2 \sigma_1^2}}-e^{-\frac{(\qeta+\mu )^2}{2 \sigma_1^2}}+r\cdot e^{-\frac{(\qeta-\mu )^2}{2 \sigma_2^2}}-r\cdot e^{-\frac{(\qeta+\mu )^2}{2 \sigma_2^2}}}
{e^{-\frac{(\qeta-\mu )^2}{2 \sigma_1^2}}+e^{-\frac{(\qeta+\mu )^2}{2 \sigma_1^2}}+r\cdot e^{-\frac{(\qeta-\mu )^2}{2 \sigma_2^2}}+r\cdot e^{-\frac{(\qeta+\mu )^2}{2 \sigma_2^2}}}.\label{doubleGaussian_A}
\end{align}
\label{eqn:doubleGaussian_SA}
\end{subequations}
\end{widetext}
\hspace*{-0.15cm}It is not clear how to simplify these. However, the inclusive $\afblep$ from Eq.~(\ref{eqn:S_A_AFB}) can be simplified to
\begin{equation}
\afblep = \frac{\sigma_1 \cdot \erf(\frac{\mu}{\sqrt{2}\sigma_1})+r\cdot\sigma_2 \cdot \erf(\frac{\mu}{\sqrt{2}\sigma_2})}{\sigma_1 + r\cdot \sigma_2}\label{eqn:afblep_DG}.
\end{equation}
This functional form is shown in Fig.~\ref{fig:AFB_mu}, and, in the limit of $\mu \ll \sigma_1$, which corresponds to $|\afblep| \lesssim 0.2$, in Fig.~\ref{fig:AFB_mu_zoomin}, we find that $\afblep = 1.22 \cdot \mu$ which approximates the data well.

\begin{figure}[p]
\begin{center}
\subfloat[\label{fig:s_qeta_mu}]{\includegraphics[width=\columnwidth]{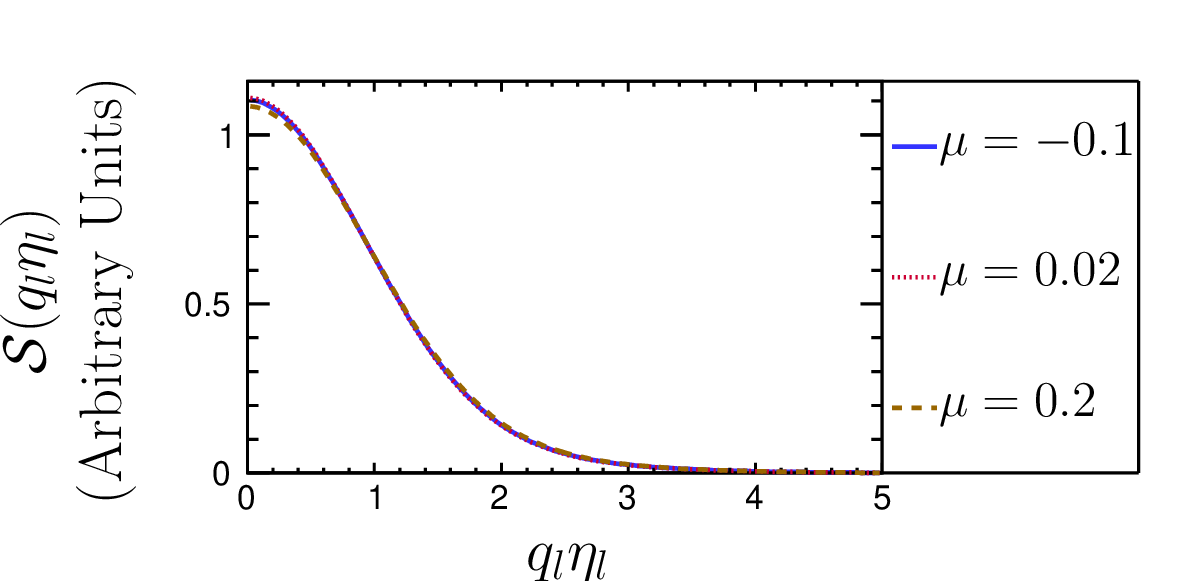}}\\
\subfloat[\label{fig:a_qeta_mu}]{\includegraphics[width=\columnwidth]{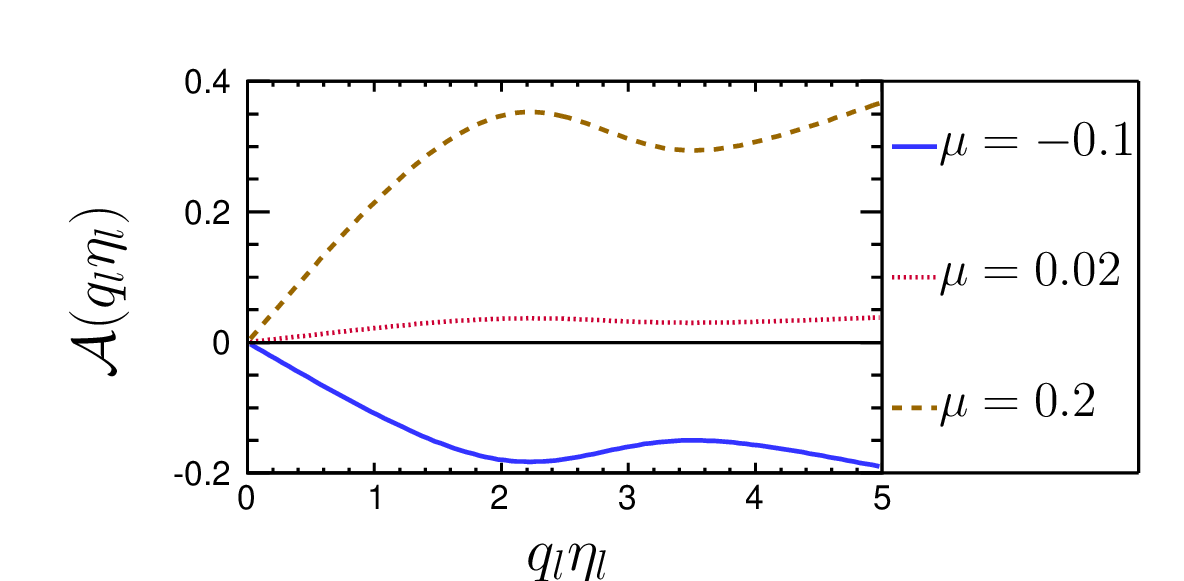}}\\
\caption{The $\Sqeta$ term and the $\Aqeta$ term from the double-Gaussian model, with the $\mu$ parameter varied.}
\label{fig:S_mu_A_mu}
\end{center}
\end{figure}

The SM prediction and most models of new physics (and the current data) all have values of $\afblep < 0.2$, so this can have a significant impact in simplifying the measurements. We can show the distribution of the $\Sqeta$ and $\Aqeta$ from Eq.~(\ref{eqn:doubleGaussian_SA}) with $\mu = -0.1, 0.02~\text{and}~0.2$ in Fig.~\ref{fig:S_mu_A_mu}. The $\Sqeta$ term is largely unchanged except for small values of $\qeta$ as previously noted, and the $\Aqeta$ term varies significantly. We also note that the distribution looks like a hyperbolic tangent function for $\qeta < 2$, but has different structures for larger values of $\qeta$.

\begin{figure}[hbtp]
\begin{center}
\subfloat[\label{fig:S_A_AFB}]{\includegraphics[width=\columnwidth]{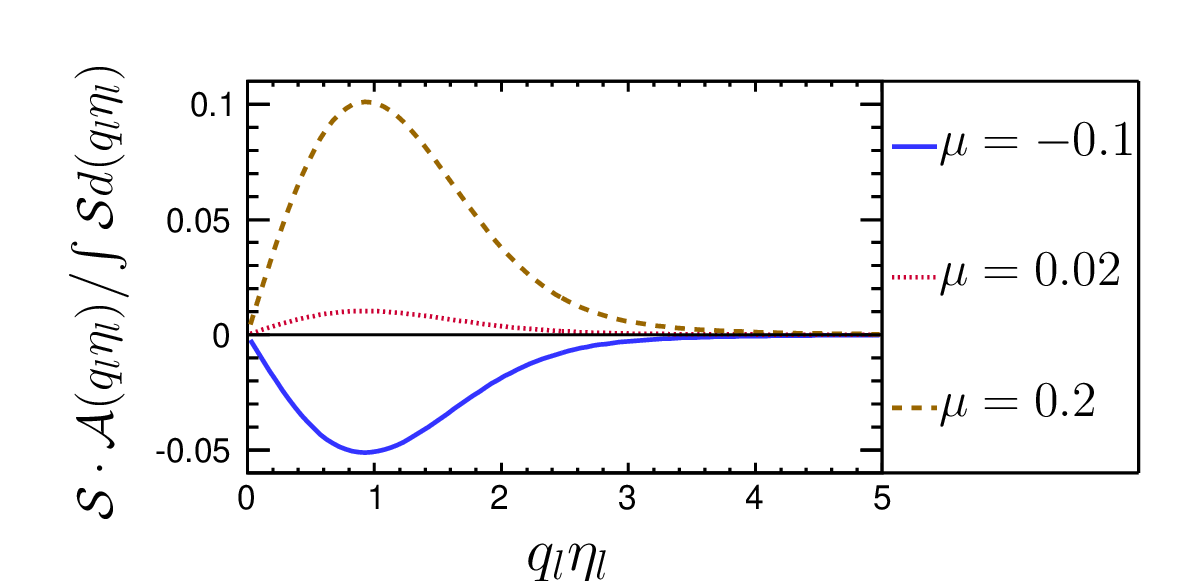}}\\
\subfloat[\label{fig:S_A_shape}]{\includegraphics[width=\columnwidth]{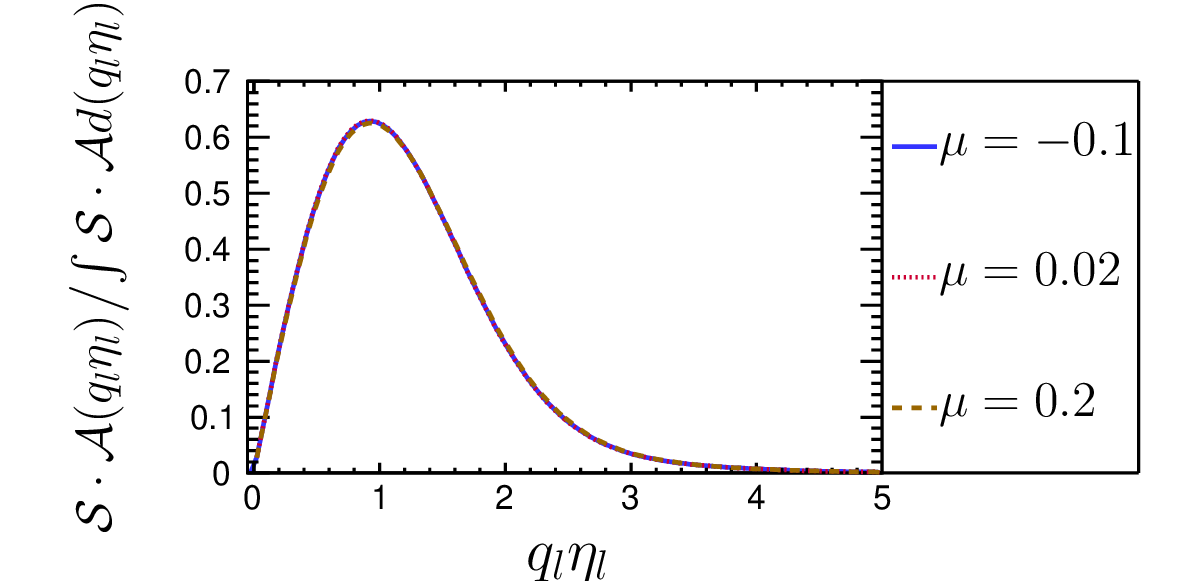}}\\
\caption{Figures showing the differential contribution to the total asymmetry as a function of $\qeta$ using the double-Gaussian model. This is estimated using the $\Sqeta$ term times the $\Aqeta$ term, with the $\mu$ parameter varied, with different overall normalizations. (a) The curves are normalized so that $\int \Sqeta~\mathrm{d}\qeta=1$ as in Eq.~(\ref{eqn:SA_Seq1}). In this case, the areas under the curves give the inclusive asymmetry. (b) The curves are normalized to $\int \Sqeta \cdot \Aqeta~ \mathrm{d}\qeta=1$ as in Eq.~(\ref{eqn:SA_SAeq1}). In this case, we can see that the differential contribution to the asymmetry as a function of $\qeta$ is largely independent of the value of $\mu$ for small values of $\mu$.}
\label{fig:S_A_AFB_shape}
\end{center}
\end{figure}

A second set of important results comes from a description of of how much contribution there is to the total asymmetry as a function of $\qeta$ (the differential contribution). It can be calculated as
\begin{equation}
\frac{\Sqeta \cdot \Aqeta}{\int_{0}^{\infty} \mathcal{S}(x)~\mathrm{d}x}\label{eqn:SA_Seq1}, 
\end{equation}
where the denominator normalizes the area under the curve to be the total asymetry. 
The results are shown in Fig.~\ref{fig:S_A_AFB} for the same three $\mu$ values. In some ways the three curves look very different, but they do share some common features. While the area under the curve is strongly dependent on $\mu$, the shape of the distribution looks remarkably similar for all three curves. To see the similarity, we plot the normalized shape by rewriting Eq.~(\ref{eqn:SA_Seq1}) such that the integral under the curve is equal to unity. Specifically:
\begin{equation}
\frac{\Sqeta \cdot \Aqeta}{\int_{0}^{\infty} \mathcal{S}(x)\cdot \mathcal{A}(x)~\mathrm{d}x}\label{eqn:SA_SAeq1}.
\end{equation}
The results are shown in Fig.~\ref{fig:S_A_shape} and we note that the shape of the differential contribution stays remarkably stable.

\begin{figure}[hbtp]
\begin{center}
\includegraphics[width=\columnwidth]{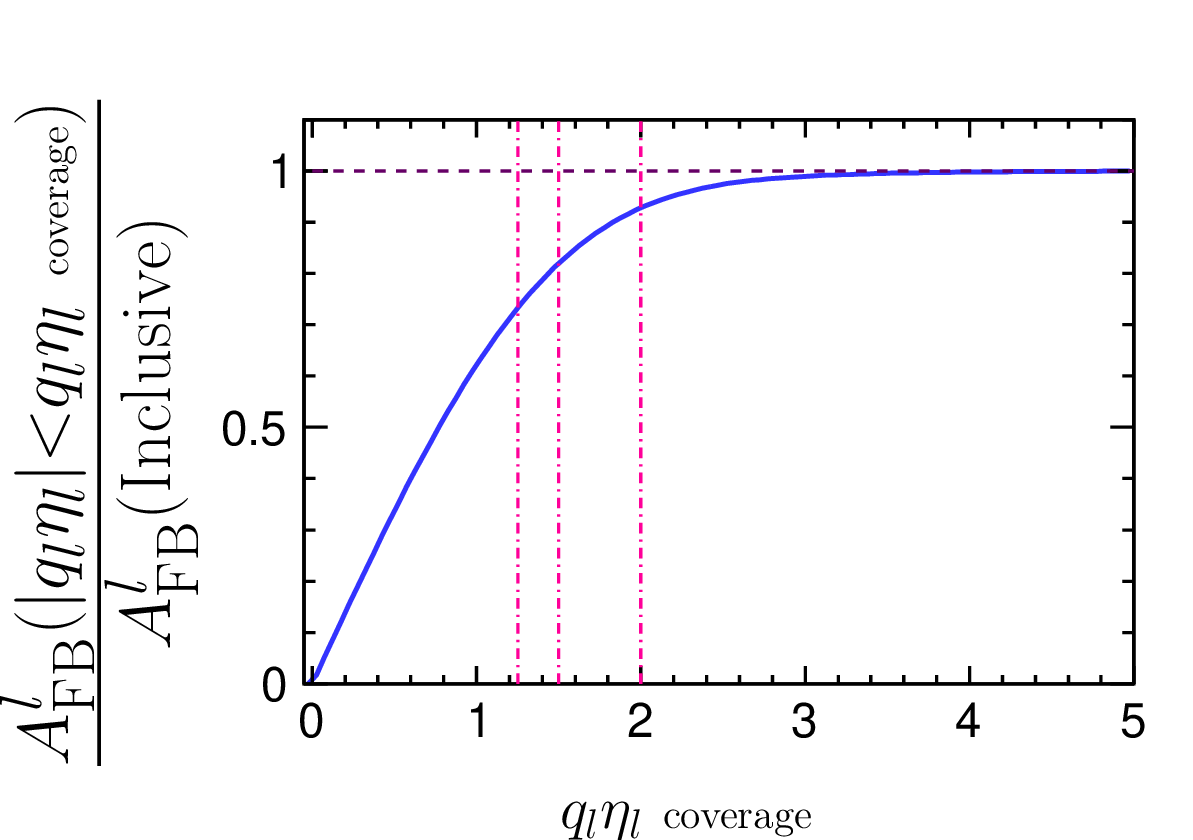}
\caption{Fraction of $\afblep$ within a certain $\qeta$ coverage. The vertical lines show $\qeta = 1.25, 1.5~\text{and}~2.0$ corresponding to the typical detector coverages at CDF and D0. The numbers are given in Table~\ref{table:AFB_fraction}. The horizontal line indicates that the fraction asymptotes to one as the $\qeta$ coverage goes to infinity.}
\label{fig:AFB_fraction}
\end{center}
\end{figure}

We are now able to make a number of further observations. First, the dominant contribution to the overall asymmetry comes from the region around $|\qeta|=1$, which is the place where the detectors have excellent coverage and resolution. We can also see why the slight mismodeling in the vicinity of $\qeta = 0$ in the $\Sqeta$ term, as shown in Fig.~\ref{fig:s_qeta_hepg}, and the mismodeling from the $\atanh$ description in the region where $\qeta > 2.5$ in the $\Aqeta$ term would only introduce small biases in the overall measurement compared to the dominant uncertainties. Specifically, even though most of the events have $|\qeta|<0.1$, the contribution to $\afblep$ from this region is $\sim$2\%. Similarly, the $\qeta$ region where there is no detector coverage at CDF or D0, $|\qeta|>2.0$, contributes $\sim$11\% to the inclusive $\afblep$; conversely, the region where the $\atanh$ fit performs poorly, $|\qeta| > 2.5$, contributes only 4\%. In addition, the constancy of the shape of the differential contribution provides an explanation for why the extrapolation technique from the measured $\afblep$ to the inclusive $\afblep$ is robust. The fraction of the $\afblep$ within certain $|\qeta|$ ranges are shown in Fig.~\ref{fig:AFB_fraction}, and some interesting numbers corresponding to typical lepton coverages at CDF and D0 are listed in Table~\ref{table:AFB_fraction}.

\begin{table}[hbtp]
\caption{Fraction of $\afblep$ within typical $\qeta$ coverage at CDF and D0.}
\begin{tabular}{cc}
\hline\hline
$\qeta$ Coverage		&	$\afblep$ Fraction\\\hline
1.25		&	0.73\\
1.5		&	0.82\\
2.0		&	0.93\\\hline\hline
\end{tabular}
\label{table:AFB_fraction}
\end{table}

\clearpage
\section{\label{sec:Comparison}Comparing the sensitivity of the $a\cdot\rm{tanh}$, single-Gaussian and double-Gaussian models}
\begin{figure}[htbp]
\begin{center}
\subfloat[\label{fig:A_models}]{\includegraphics[width=\columnwidth]{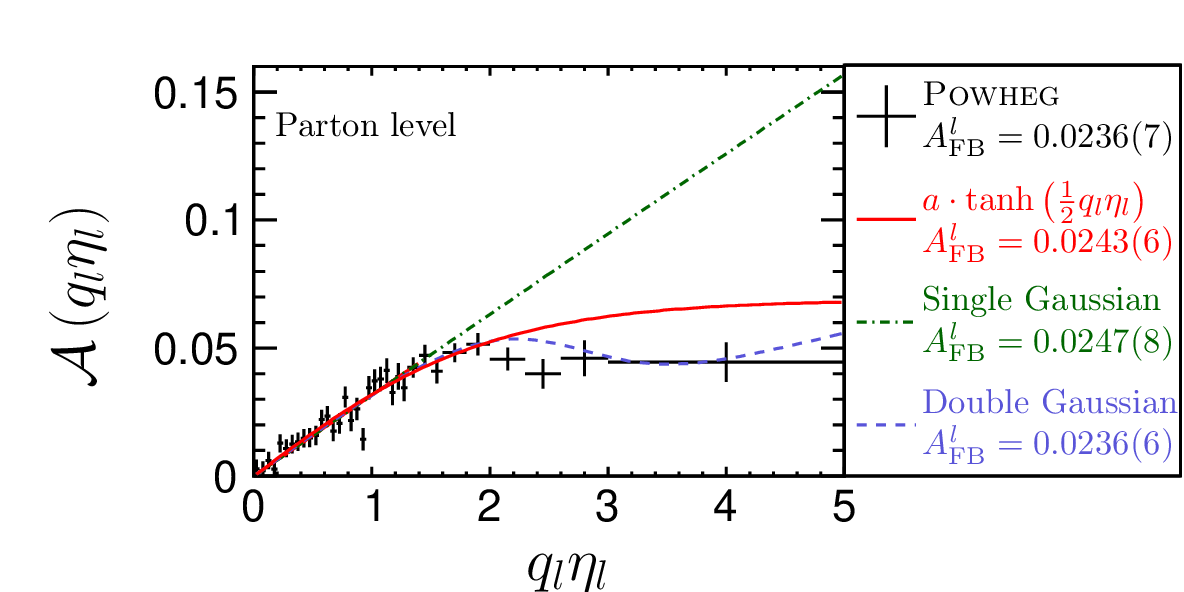}}\\
\subfloat[\label{fig:S_A_models}]{\includegraphics[width=\columnwidth]{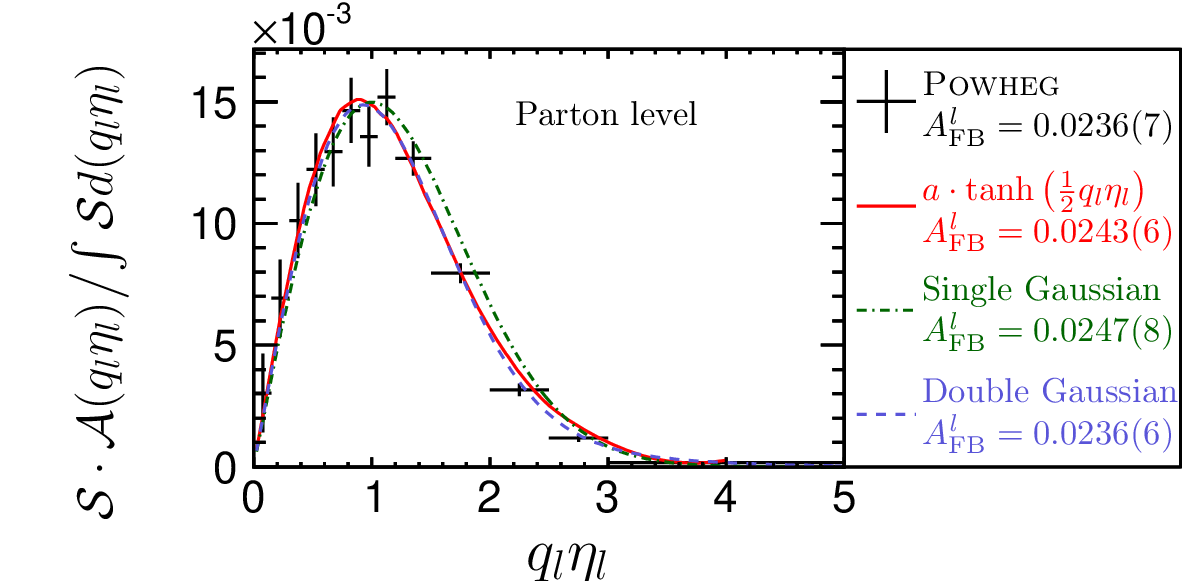}}\\
\caption{Comparison among the $\atanh$ model, the single-Gaussian model, the double-Gaussian model and the \textsc{powheg} simulation. (a) shows the best fits of the $\Aqeta$ distribution (done only using events with $|\qeta|<2.0$), while (b) shows the differential contribution to the $\afblep$ as a function of $\qeta$ from different models.}
\label{fig:A_S_A_models}
\end{center}
\end{figure}

\begin{figure}[htbp]
\begin{center}
\includegraphics[width=\columnwidth]{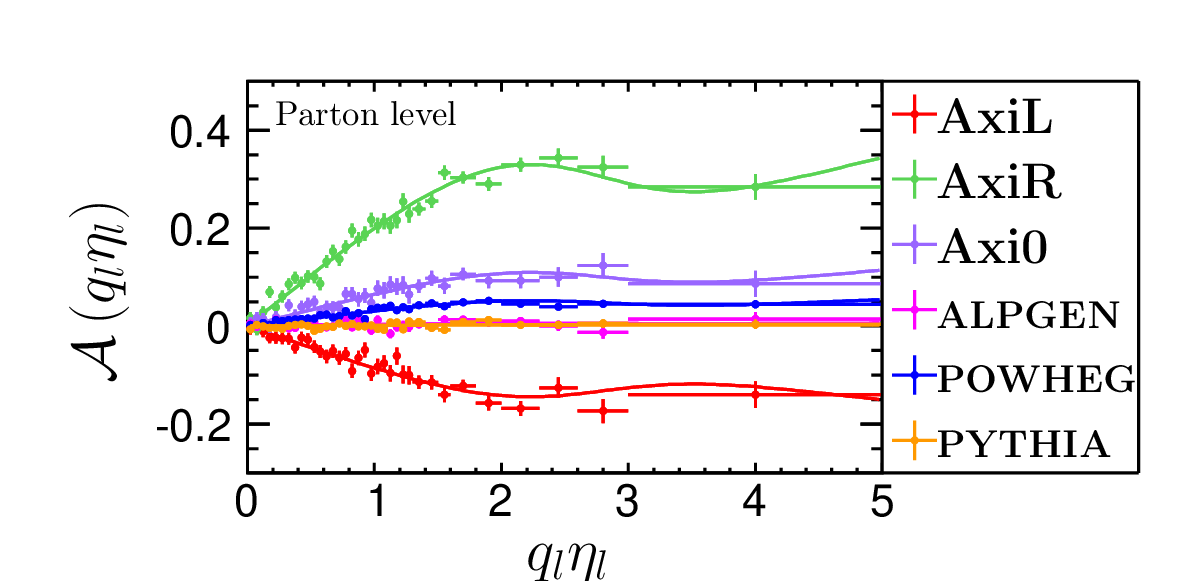}\\
\end{center}
\caption{Best fit of double-Gaussian model to the $\Aqeta$ distribution for various $\ttbar$ samples at generator level. This figure can be compared directly to Fig.~\ref{fig:a_qeta_hepg} where we fit the same data, but using the $\atanh$ function.}
\label{fig:A_hepg_6samples}
\end{figure}

We compare the sensitivity of the possible measurement techniques in a number of ways. First we compare them visually, then we consider how well the different measurement techniques would work. Fig~\ref{fig:A_S_A_models} shows the $\Aqeta$ term and the differential contribution to the inclusive $\afblep$ as a function of $\qeta$ from the \textsc{powheg} sample, overlaid with the best fit from the $\atanh$ model, the single-Gaussian model and the double-Gaussian model described in this article, when we only consider events with $|\qeta|<2.0$. All three models fit this $\qeta$ region well. Since the region $|\qeta|<2.0$  is where most of the contribution to $\afblep$ comes from, all three models (including the single-Gaussian model) get back to the inclusive $\afblep$ of the sample reasonably well. The double-Gaussian model fits the asymmetric part better in the $\qeta$ region above 2.0 than the tanh model, thus the differential contribution predicted by the double-Gaussian model lines up with the \textsc{powheg} predicted points marginally better. However, as stated earlier, the improvement is in the region where the contribution to the inclusive $\afblep$ is small, thus the improvement in the resultant $\afblep$ using the double-Gaussian model is very small. Fig.~\ref{fig:A_hepg_6samples} shows the double-Gaussian model fit to the $\Aqeta$ distribution for all the six benchmark samples at parton level. A comparison with Fig.~\ref{fig:a_qeta_hepg} shows that the double-Gaussian model matches all the simulated samples better than the $a\cdot$tanh model, although the differences are mostly in the high-$\qeta$ region where the contribution to the inclusive $\afblep$ is small, and there is no data from the experiments in this region.

We next compare how well the various methods will work for real data by considering just the set of \textsc{powheg} simulated events within $|\qeta| < 2.0$ and employing different methodologies to see how well each reproduces the inclusive $\afblep$ of 0.0236. We performed 10000 pseudo-experiments by varying the $\mathrm{d}\mathcal{N}(\qeta)/\mathrm{d}(\qeta)$ distribution with statistical fluctuations for about 1 million simulated events. We then measured $\afblep$ for each pseudo-experiment using each of the four methods:
\begin{enumerate}
\item A pure counting of the number of events with positive and negative $\qeta$ values, with a correction for the limited detector coverage using the correction factor of 0.93 (see Table~\ref{table:AFB_fraction}) to extrapolate to the inclusive value
\item Fitting the $\atanh$ model to the $\Aqeta$ term of the distribution for the parameter $a$ and calculating the inclusive $\afblep$ using the $\Sqeta$ distribution and Eq.~(\ref{eqn:S_A_AFB})
\item Fitting the asymmetric part of the double-Gaussian model to the $\Aqeta$ term of the distribution for the parameter $\mu$ and calculating the inclusive $\afblep$ with Eq.~(\ref{eqn:afblep_DG})
\item Fitting the double-Gaussian model to the $\qeta$ distribution itself for the parameter $\mu$ and calculating the inclusive $\afblep$ again with Eq.~(\ref{eqn:afblep_DG})
\end{enumerate}
The results of the pseudo-experiments are shown in Table~\ref{table:PE4Methods}.

The average of the pseudo-experiments for each method is always within one standard deviation of the input $\afblep$ value, indicating none have noticeable bias. As expected, the pure counting method has the largest uncertainty, as the fits incorporate the additional shape information to reduce the uncertainties. While there does not seem to be much difference in the sensitivity of the fitting methods, we note that the fit on the $\Aqeta$ term has the systematic advantage over the pure fit to the mean, $\mu$, of the full $\mathrm{d}\mathcal{N}(\qeta)/\mathrm{d}(\qeta)$ distribution as most of the systematic uncertainties due to the acceptance of the detector are expected to cancel out~\cite{Aaltonen:2013vaf}. Thus, we favor the use of the fit on the $\Aqeta$ distribution over the simple counting for resolution reasons, and over the fit on the full distribution for robustness reasons. Between the two fits on the $\Aqeta$ term, we see that the $\atanh$ formulation is easier to work with, but further checks to see if there are other effects due to detector response should be considered.

\begin{table*}[hbtp]
\caption{Results of pseudo-experiments using the different methods to reproduce $\afblep$ of the \textsc{powheg} simulation (0.0236), but only using events with $|\qeta| < 2.0$. Note that the uncertainties listed are statistical only and are due to the size of the simulated data sample.}
\begin{tabular}{c c c c}
\hline
Method						&	Mean		&	Mean-Expected	& 	Uncertainty \\\hline\hline
Counting						&	0.0241	&	0.0004			&	0.0008		\\
$\atanh$ $\Aqeta$ Fit		& 	0.0243	&	0.0006 			& 	0.0006\\
Double Gaussian $\Aqeta$ Fit	&	0.0236	&	-0.0001 			& 	0.0006\\
Double Gaussian Direct Fit 	&	0.0238	&	0.0002 			& 	0.0006\\\hline
\end{tabular}
\label{table:PE4Methods}
\end{table*}

We next test how well the $\atanh$ and the double-Gaussian methods reproduce the inclusive $\afblep$ values for all 6 simulated samples with only events within $|\qeta|<2.0$. A comparison of results is given in Table~\ref{table:AFB_models_gen}. Though the double-Gaussian model works better in the high $\qeta$ region, the impact on the $\afblep$ measurement is negligible compared to the dominant uncertainties in the measurement ($\sim$0.02 in the lepton+jets final state~\cite{Aaltonen:2013vaf} and $\sim$0.05 in the dilepton final state~\cite{CDF11035}). 

\begin{table*}[htbp]
        \caption{Comparison of the predicted $\afblep$ values and the corresponding measured $\afblep$ values with the $\atanh$ model and the double-Gaussian model. The uncertainties are statistical only and are always small compared to the expected statistical uncertainty in data collected by the CDF and D0 experiments.}
\begin{tabular}{cccc}
        \hline
        \multirow{2}{*}{Model} & \multirow{2}{*}{True $\afblep$}& Measured $\afblep$& Measured $\afblep$\\
        & & ($\atanh$ model)& (Double-Gaussian model)\\
        \hline
        \hline
        AxiL & -0.063(2)  & -0.064(2) & -0.064(2) \\
        AxiR & 0.151(2)   & 0.148(2)  & 0.150(2)  \\
        Axi0 & 0.050(2)   & 0.048(2)  & 0.048(2) \\
        \textsc{Alpgen} & 0.003(1) &-0.004(1)  & 0.002(1)  \\
        \textsc{Pythia} & 0.001(1) &-0.005(1)  & 0.001(1)  \\
        \textsc{powheg} &0.023(1)  & 0.024(1)  & 0.023(1) \\
        \hline
        \end{tabular}
        \label{table:AFB_models_gen}
\end{table*}

Finally, on a related measurement note we point out that, because of the predicted structure in the high-$\qeta$ region, when computing $\Aqeta$ the choice of the $\qeta$ bin centroids and widths should be made with care. Each bin should contain a reasonable number of events to avoid statistical fluctuations; on the other hand, as shown in Fig.~\ref{fig:A_hepg_6samples}, the curve is changing drastically above $\qeta \sim 1.5$, thus a simple fit through either the bin center or the bin centroid could introduce a sizeable systematic uncertainty if the bin is overly wide.

\section{\label{sec:Conclusion}Conclusion}
We have described the $\qeta$ distribution from the leptonic cascade decays of $\ttbar$ events produced at the Tevatron and the corresponding $\afblep$ that can be determined from it. Many data measurements have been produced in this final state, all of which only have coverage of $|\eta|<2.0$, and some have used a empirical functional form of $\atanh$ to extrapolate from the limited detector $\eta$ coverage to an inclusive parton-level estimate. We now understand that this excellent approximation is fortuitous but robust. The $\atanh$ parametrization is an approximation that is only good for values of $|\qeta|<2.5$, but it is more than good enough for the Tevatron experiments. It may well be useful for all Tevatron experiments to report their $\afblep$ in the restricted $\qeta$ regimes as well as measurements of the $a$ term in the $\atanh$ formulation if possible.

Our studies show that a more sophisticated empirical function, which takes the form of the sum of two Gaussian functions with a common mean, and with empirically determined values of the two $\sigma$ and $r$ parameters, describes the $\qeta$ distribution better at all $\qeta$ values. This functional form has not yielded a simple closed form for the $\Aqeta$ term. While the double-Gaussian parametrization is better in principle, in practice using it does not provide additional useful measurement sensitivity and it is more cumbersome to use. On the other hand, this better understanding of the expected shapes lead to some interesting and useful conclusions in addition to the confidence we now have in the methods previously being employed. First, it is advantageous to think of the asymmetry as coming from the shift of the mean of the $\qeta$ distribution. To a good degree of approximation, measuring the $\afblep$ is equivalent to measuring the mean, $\mu$, in the limit of small $\afblep$; measuring the $\Aqeta$ term of the distribution is one of a number of ways to do so, which also takes advantage of the cancelling of the systematic uncertainties caused by the detector response in the measurement. Ultimately, we now understand that the dominant contribution to the inclusive $\afblep$ comes from the region within the CDF and D0 detectors which are best covered, and that the extrapolation procedures allow for a robust measurement.

%

\section{\label{sec:Acknowledgements}Acknowledgements}
The authors would like to thank Dr. Michelangelo Mangano for the discussion about the origin of the double-Gaussian model. We also would like to thank Hamilton Carter and Dr.~Ilarion Melnikov for the useful discussions about turning the integrations of Gaussian distributions into hyperbolic tangent functions. We would like to thank FNAL and the CDF collaboration for their support while this work was done. SH, ZH and DT would also like to thank the Mitchell Institute for Fundamental Physics and Astronomy and the Department of Physics and Astronomy at Texas A\&M University for its support.

\clearpage
\bibliographystyle{apsrev4-1-JHEPfix}
\bibliography{PubCitations}

\end{document}

%% file: afbcommands.tex
\newcommand{\MetVec}{\mbox{$\vec E\kern-0.50em\raise0.10ex\hbox{/}_{T}$}}
\newcommand{\MetVecRaw}{\mbox{$\vec E\kern-0.50em\raise0.10ex\hbox{/}^{raw}_{T}$}}
\newcommand{\Met}{\mbox{$E\kern-0.50em\raise0.10ex\hbox{/}_{T}$}}
\newcommand{\MetRaw}{\mbox{$E\kern-0.50em\raise0.10ex\hbox{/}_{T}^{raw}$}}
\newcommand{\MetSpec}{\mbox{$E\kern-0.50em\raise0.10ex\hbox{/}_{T}^{spec}$}}

\newcommand{\MetSig}{\mbox{$E\kern-0.50em\raise0.10ex\hbox{/}_{T}^{sig}$}}

\newcommand{\ttbar}{t \bar{t}}

\newcommand{\gevcc}{\rm{GeV}/c^2}

\newcommand{\MET}{\mbox{$E\kern-0.50em\raise0.10ex\hbox{/}_{T}$}}
\newcommand{\met}{\mbox{$E\kern-0.50em\raise0.10ex\hbox{/}_{T}$}}
\newcommand{\METzero}{\mbox{$E\kern-0.50em\raise0.10ex\hbox{/}_{T}^0$}}
\newcommand{\vecMET}{\mbox{$\vec{E}\kern-0.50em\raise0.10ex\hbox{/}_{T}$}}
\newcommand{\METSPEC}{\mbox{$E\kern-0.50em\raise0.10ex\hbox{/}_{Tspec}$}}

\def\slantfrac#1#2{\kern.1em^{#1}\kern-.3em/\kern-.1em_{#2}}

\newcommand{\afb}{A_{\text{FB}}}

\newcommand{\qeta}{q_{l}\eta_{l}}

\newcommand{\afblep}{A_{\text{FB}}^{\it{l}}}

\DeclareUnicodeCharacter{2212}{\textminus}
\mathchardef\mhyphen="2D
\newcommand{\erf}{\mathrm{erf}\,}

\newcommand{\Sqeta}{\mathcal{S}(\qeta)}
\newcommand{\Aqeta}{\mathcal{A}(\qeta)}

\newcommand{\atanh}{a \cdot \rm{tanh}}
\newcommand{\afbtt}{A_{\text{FB}}^{t\bar{t}}}

%% file: DoubleGaussPRD.bbl
\begin{thebibliography}{43}%
\makeatletter
\providecommand \@ifxundefined [1]{%
 \@ifx{#1\undefined}
}%
\providecommand \@ifnum [1]{%
 \ifnum #1\expandafter \@firstoftwo
 \else \expandafter \@secondoftwo
 \fi
}%
\providecommand \@ifx [1]{%
 \ifx #1\expandafter \@firstoftwo
 \else \expandafter \@secondoftwo
 \fi
}%
\providecommand \natexlab [1]{#1}%
\providecommand \enquote  [1]{``#1''}%
\providecommand \bibnamefont  [1]{#1}%
\providecommand \bibfnamefont [1]{#1}%
\providecommand \citenamefont [1]{#1}%
\providecommand \href@noop [0]{\@secondoftwo}%
\providecommand \href [0]{\begingroup \@sanitize@url \@href}%
\providecommand \@href[1]{\@@startlink{#1}\@@href}%
\providecommand \@@href[1]{\endgroup#1\@@endlink}%
\providecommand \@sanitize@url [0]{\catcode `\\12\catcode `\$12\catcode
  `\&12\catcode `\#12\catcode `\^12\catcode `\_12\catcode `\%12\relax}%
\providecommand \@@startlink[1]{}%
\providecommand \@@endlink[0]{}%
\providecommand \url  [0]{\begingroup\@sanitize@url \@url }%
\providecommand \@url [1]{\endgroup\@href {#1}{\urlprefix }}%
\providecommand \urlprefix  [0]{URL }%
\providecommand \Eprint [0]{\href }%
\providecommand \doibase [0]{http://dx.doi.org/}%
\providecommand \selectlanguage [0]{\@gobble}%
\providecommand \bibinfo  [0]{\@secondoftwo}%
\providecommand \bibfield  [0]{\@secondoftwo}%
\providecommand \translation [1]{[#1]}%
\providecommand \BibitemOpen [0]{}%
\providecommand \bibitemStop [0]{}%
\providecommand \bibitemNoStop [0]{.\EOS\space}%
\providecommand \EOS [0]{\spacefactor3000\relax}%
\providecommand \BibitemShut  [1]{\csname bibitem#1\endcsname}%
\let\auto@bib@innerbib\@empty
\bibitem [{\citenamefont {Aaltonen}\ \emph
  {et~al.}(2013{\natexlab{a}})\citenamefont {Aaltonen} \emph
  {et~al.}}]{Aaltonen:2012it}%
  \BibitemOpen
  \bibfield  {author} {\bibinfo {author} {\bibfnamefont {T.}~\bibnamefont
  {Aaltonen}}\  \emph {et~al.} (\bibinfo {collaboration} {CDF Collaboration}),\
  }\href@noop {} {\bibfield  {journal} {\bibinfo  {journal} {Phys. Rev. D}\
  }\textbf {\bibinfo {volume} {87}},\ \bibinfo {pages} {092002} (\bibinfo
  {year} {2013}{\natexlab{a}})}\BibitemShut {NoStop}%
\bibitem [{\citenamefont {Abazov}\ \emph {et~al.}(2011)\citenamefont {Abazov}
  \emph {et~al.}}]{d0_afb_prd2011}%
  \BibitemOpen
  \bibfield  {author} {\bibinfo {author} {\bibfnamefont {V.}~\bibnamefont
  {Abazov}}\  \emph {et~al.} (\bibinfo {collaboration} {D0 Collaboration}),\
  }\href@noop {} {\bibfield  {journal} {\bibinfo  {journal} {Phys. Rev. D}\
  }\textbf {\bibinfo {volume} {84}},\ \bibinfo {pages} {112005} (\bibinfo
  {year} {2011})}\BibitemShut {NoStop}%
\bibitem [{\citenamefont {Aaltonen}\ \emph
  {et~al.}(2013{\natexlab{b}})\citenamefont {Aaltonen} \emph
  {et~al.}}]{CDF:2013gna}%
  \BibitemOpen
  \bibfield  {author} {\bibinfo {author} {\bibfnamefont {T.}~\bibnamefont
  {Aaltonen}}\  \emph {et~al.} (\bibinfo {collaboration} {CDF Collaboration}),\
  }\href@noop {} {\bibfield  {journal} {\bibinfo  {journal} {Phys. Rev. Lett.}\
  }\textbf {\bibinfo {volume} {111}},\ \bibinfo {pages} {182002} (\bibinfo
  {year} {2013}{\natexlab{b}})}\BibitemShut {NoStop}%
\bibitem [{\citenamefont {Bernreuther}\ and\ \citenamefont
  {Si}(2012)}]{Bernreuther:2012sx}%
  \BibitemOpen
  \bibfield  {author} {\bibinfo {author} {\bibfnamefont {W.}~\bibnamefont
  {Bernreuther}}\ and\ \bibinfo {author} {\bibfnamefont {Z.-G.}\ \bibnamefont
  {Si}},\ }\href@noop {} {\bibfield  {journal} {\bibinfo  {journal} {Phys. Rev.
  D}\ }\textbf {\bibinfo {volume} {86}},\ \bibinfo {pages} {034026} (\bibinfo
  {year} {2012})}\BibitemShut {NoStop}%
\bibitem [{\citenamefont {K\"uhn}\ and\ \citenamefont
  {Rodrigo}(1998)}]{PhysRevLett.81.49}%
  \BibitemOpen
  \bibfield  {author} {\bibinfo {author} {\bibfnamefont {J.~H.}\ \bibnamefont
  {K\"uhn}}\ and\ \bibinfo {author} {\bibfnamefont {G.}~\bibnamefont
  {Rodrigo}},\ }\href@noop {} {\bibfield  {journal} {\bibinfo  {journal} {Phys.
  Rev. Lett.}\ }\textbf {\bibinfo {volume} {81}},\ \bibinfo {pages} {49}
  (\bibinfo {year} {1998})}\BibitemShut {NoStop}%
\bibitem [{\citenamefont {Jung}\ \emph {et~al.}(2011)\citenamefont {Jung},
  \citenamefont {Ko},\ and\ \citenamefont {Lee}}]{Jung:2010yn}%
  \BibitemOpen
  \bibfield  {author} {\bibinfo {author} {\bibfnamefont {D.-W.}\ \bibnamefont
  {Jung}}, \bibinfo {author} {\bibfnamefont {P.}~\bibnamefont {Ko}}, and\
  \bibinfo {author} {\bibfnamefont {J.~S.}\ \bibnamefont {Lee}},\ }\href@noop
  {} {\bibfield  {journal} {\bibinfo  {journal} {Phys. Lett. B}\ }\textbf
  {\bibinfo {volume} {701}},\ \bibinfo {pages} {248} (\bibinfo {year}
  {2011})}\BibitemShut {NoStop}%
\bibitem [{\citenamefont {Jung}\ \emph
  {et~al.}(2010{\natexlab{a}})\citenamefont {Jung}, \citenamefont {Ko},
  \citenamefont {Lee},\ and\ \citenamefont {hyeon Nam}}]{Jung2010238}%
  \BibitemOpen
  \bibfield  {author} {\bibinfo {author} {\bibfnamefont {D.-W.}\ \bibnamefont
  {Jung}}, \bibinfo {author} {\bibfnamefont {P.}~\bibnamefont {Ko}}, \bibinfo
  {author} {\bibfnamefont {J.~S.}\ \bibnamefont {Lee}}, and\ \bibinfo {author}
  {\bibfnamefont {S.}~\bibnamefont {hyeon Nam}},\ }\href@noop {} {\bibfield
  {journal} {\bibinfo  {journal} {Phys. Lett. B}\ }\textbf {\bibinfo {volume}
  {691}},\ \bibinfo {pages} {238 } (\bibinfo {year}
  {2010}{\natexlab{a}})}\BibitemShut {NoStop}%
\bibitem [{\citenamefont {Frampton}\ \emph {et~al.}(2010)\citenamefont
  {Frampton}, \citenamefont {Shu},\ and\ \citenamefont
  {Wang}}]{Frampton2010294}%
  \BibitemOpen
  \bibfield  {author} {\bibinfo {author} {\bibfnamefont {P.~H.}\ \bibnamefont
  {Frampton}}, \bibinfo {author} {\bibfnamefont {J.}~\bibnamefont {Shu}}, and\
  \bibinfo {author} {\bibfnamefont {K.}~\bibnamefont {Wang}},\ }\href@noop {}
  {\bibfield  {journal} {\bibinfo  {journal} {Phys. Lett. B}\ }\textbf
  {\bibinfo {volume} {683}},\ \bibinfo {pages} {294 } (\bibinfo {year}
  {2010})}\BibitemShut {NoStop}%
\bibitem [{\citenamefont {Álvarez}\ \emph {et~al.}(2011)\citenamefont
  {Álvarez}, \citenamefont {Rold},\ and\ \citenamefont
  {Szynkman}}]{jhep05(2011)070}%
  \BibitemOpen
  \bibfield  {author} {\bibinfo {author} {\bibfnamefont {E.}~\bibnamefont
  {Álvarez}}, \bibinfo {author} {\bibfnamefont {L.}~\bibnamefont {Rold}}, and\
  \bibinfo {author} {\bibfnamefont {A.}~\bibnamefont {Szynkman}},\ }\href@noop
  {} {\bibfield  {journal} {\bibinfo  {journal} {J. High Energy Phys.}\
  }\bibinfo {volume} {05} (\bibinfo {year} {2011})\ \bibinfo {pages}
  {070}}\BibitemShut {NoStop}%
\bibitem [{\citenamefont {Chen}\ \emph {et~al.}(2011)\citenamefont {Chen},
  \citenamefont {Cvetic},\ and\ \citenamefont {Kim}}]{Chen:2010hm}%
  \BibitemOpen
  \bibfield  {author} {\bibinfo {author} {\bibfnamefont {C.-H.}\ \bibnamefont
  {Chen}}, \bibinfo {author} {\bibfnamefont {G.}~\bibnamefont {Cvetic}}, and\
  \bibinfo {author} {\bibfnamefont {C.}~\bibnamefont {Kim}},\ }\href@noop {}
  {\bibfield  {journal} {\bibinfo  {journal} {Phys. Lett. B}\ }\textbf
  {\bibinfo {volume} {694}},\ \bibinfo {pages} {393} (\bibinfo {year}
  {2011})}\BibitemShut {NoStop}%
\bibitem [{\citenamefont {Wang}\ \emph {et~al.}(2010)\citenamefont {Wang},
  \citenamefont {Xiao},\ and\ \citenamefont {Zhu}}]{PhysRevD.82.094011}%
  \BibitemOpen
  \bibfield  {author} {\bibinfo {author} {\bibfnamefont {Y.-k.}\ \bibnamefont
  {Wang}}, \bibinfo {author} {\bibfnamefont {B.}~\bibnamefont {Xiao}}, and\
  \bibinfo {author} {\bibfnamefont {S.-h.}\ \bibnamefont {Zhu}},\ }\href@noop
  {} {\bibfield  {journal} {\bibinfo  {journal} {Phys. Rev. D}\ }\textbf
  {\bibinfo {volume} {82}},\ \bibinfo {pages} {094011} (\bibinfo {year}
  {2010})}\BibitemShut {NoStop}%
\bibitem [{\citenamefont {Djouadi}\ \emph {et~al.}(2010)\citenamefont
  {Djouadi}, \citenamefont {Moreau}, \citenamefont {Richard},\ and\
  \citenamefont {Singh}}]{PhysRevD.82.071702}%
  \BibitemOpen
  \bibfield  {author} {\bibinfo {author} {\bibfnamefont {A.}~\bibnamefont
  {Djouadi}}, \bibinfo {author} {\bibfnamefont {G.}~\bibnamefont {Moreau}},
  \bibinfo {author} {\bibfnamefont {F.}~\bibnamefont {Richard}}, and\ \bibinfo
  {author} {\bibfnamefont {R.~K.}\ \bibnamefont {Singh}},\ }\href@noop {}
  {\bibfield  {journal} {\bibinfo  {journal} {Phys. Rev. D}\ }\textbf {\bibinfo
  {volume} {82}},\ \bibinfo {pages} {071702} (\bibinfo {year}
  {2010})}\BibitemShut {NoStop}%
\bibitem [{\citenamefont {Chivukula}\ \emph {et~al.}(2010)\citenamefont
  {Chivukula}, \citenamefont {Simmons},\ and\ \citenamefont
  {Yuan}}]{PhysRevD.82.094009}%
  \BibitemOpen
  \bibfield  {author} {\bibinfo {author} {\bibfnamefont {R.~S.}\ \bibnamefont
  {Chivukula}}, \bibinfo {author} {\bibfnamefont {E.~H.}\ \bibnamefont
  {Simmons}}, and\ \bibinfo {author} {\bibfnamefont {C.-P.}\ \bibnamefont
  {Yuan}},\ }\href@noop {} {\bibfield  {journal} {\bibinfo  {journal} {Phys.
  Rev. D}\ }\textbf {\bibinfo {volume} {82}},\ \bibinfo {pages} {094009}
  (\bibinfo {year} {2010})}\BibitemShut {NoStop}%
\bibitem [{\citenamefont {Xiao}\ \emph {et~al.}(2010)\citenamefont {Xiao},
  \citenamefont {Wang},\ and\ \citenamefont {Zhu}}]{PhysRevD.82.034026}%
  \BibitemOpen
  \bibfield  {author} {\bibinfo {author} {\bibfnamefont {B.}~\bibnamefont
  {Xiao}}, \bibinfo {author} {\bibfnamefont {Y.-k.}\ \bibnamefont {Wang}}, and\
  \bibinfo {author} {\bibfnamefont {S.-h.}\ \bibnamefont {Zhu}},\ }\href@noop
  {} {\bibfield  {journal} {\bibinfo  {journal} {Phys. Rev. D}\ }\textbf
  {\bibinfo {volume} {82}},\ \bibinfo {pages} {034026} (\bibinfo {year}
  {2010})}\BibitemShut {NoStop}%
\bibitem [{\citenamefont {Cao}\ \emph {et~al.}(2010{\natexlab{a}})\citenamefont
  {Cao}, \citenamefont {McKeen}, \citenamefont {Rosner}, \citenamefont
  {Shaughnessy},\ and\ \citenamefont {Wagner}}]{PhysRevD.81.114004}%
  \BibitemOpen
  \bibfield  {author} {\bibinfo {author} {\bibfnamefont {Q.-H.}\ \bibnamefont
  {Cao}}, \bibinfo {author} {\bibfnamefont {D.}~\bibnamefont {McKeen}},
  \bibinfo {author} {\bibfnamefont {J.~L.}\ \bibnamefont {Rosner}}, \bibinfo
  {author} {\bibfnamefont {G.}~\bibnamefont {Shaughnessy}}, and\ \bibinfo
  {author} {\bibfnamefont {C.~E.~M.}\ \bibnamefont {Wagner}},\ }\href@noop {}
  {\bibfield  {journal} {\bibinfo  {journal} {Phys. Rev. D}\ }\textbf {\bibinfo
  {volume} {81}},\ \bibinfo {pages} {114004} (\bibinfo {year}
  {2010}{\natexlab{a}})}\BibitemShut {NoStop}%
\bibitem [{\citenamefont {Dor\ifmmode~\check{s}\else \v{s}\fi{}ner}\ \emph
  {et~al.}(2010)\citenamefont {Dor\ifmmode~\check{s}\else \v{s}\fi{}ner},
  \citenamefont {Fajfer}, \citenamefont {Kamenik},\ and\ \citenamefont
  {Ko\ifmmode~\check{s}\else \v{s}\fi{}nik}}]{PhysRevD.81.055009}%
  \BibitemOpen
  \bibfield  {author} {\bibinfo {author} {\bibfnamefont {I.}~\bibnamefont
  {Dor\ifmmode~\check{s}\else \v{s}\fi{}ner}}, \bibinfo {author} {\bibfnamefont
  {S.}~\bibnamefont {Fajfer}}, \bibinfo {author} {\bibfnamefont {J.~F.}\
  \bibnamefont {Kamenik}}, and\ \bibinfo {author} {\bibfnamefont
  {N.}~\bibnamefont {Ko\ifmmode~\check{s}\else \v{s}\fi{}nik}},\ }\href@noop {}
  {\bibfield  {journal} {\bibinfo  {journal} {Phys. Rev. D}\ }\textbf {\bibinfo
  {volume} {81}},\ \bibinfo {pages} {055009} (\bibinfo {year}
  {2010})}\BibitemShut {NoStop}%
\bibitem [{\citenamefont {Jung}\ \emph
  {et~al.}(2010{\natexlab{b}})\citenamefont {Jung}, \citenamefont {Murayama},
  \citenamefont {Pierce},\ and\ \citenamefont {Wells}}]{PhysRevD.81.015004}%
  \BibitemOpen
  \bibfield  {author} {\bibinfo {author} {\bibfnamefont {S.}~\bibnamefont
  {Jung}}, \bibinfo {author} {\bibfnamefont {H.}~\bibnamefont {Murayama}},
  \bibinfo {author} {\bibfnamefont {A.}~\bibnamefont {Pierce}}, and\ \bibinfo
  {author} {\bibfnamefont {J.~D.}\ \bibnamefont {Wells}},\ }\href@noop {}
  {\bibfield  {journal} {\bibinfo  {journal} {Phys. Rev. D}\ }\textbf {\bibinfo
  {volume} {81}},\ \bibinfo {pages} {015004} (\bibinfo {year}
  {2010}{\natexlab{b}})}\BibitemShut {NoStop}%
\bibitem [{\citenamefont {Shu}\ \emph {et~al.}(2010)\citenamefont {Shu},
  \citenamefont {Tait},\ and\ \citenamefont {Wang}}]{PhysRevD.81.034012}%
  \BibitemOpen
  \bibfield  {author} {\bibinfo {author} {\bibfnamefont {J.}~\bibnamefont
  {Shu}}, \bibinfo {author} {\bibfnamefont {T.~M.~P.}\ \bibnamefont {Tait}},
  and\ \bibinfo {author} {\bibfnamefont {K.}~\bibnamefont {Wang}},\ }\href@noop
  {} {\bibfield  {journal} {\bibinfo  {journal} {Phys. Rev. D}\ }\textbf
  {\bibinfo {volume} {81}},\ \bibinfo {pages} {034012} (\bibinfo {year}
  {2010})}\BibitemShut {NoStop}%
\bibitem [{\citenamefont {Arhrib}\ \emph {et~al.}(2010)\citenamefont {Arhrib},
  \citenamefont {Benbrik},\ and\ \citenamefont {Chen}}]{PhysRevD.82.034034}%
  \BibitemOpen
  \bibfield  {author} {\bibinfo {author} {\bibfnamefont {A.}~\bibnamefont
  {Arhrib}}, \bibinfo {author} {\bibfnamefont {R.}~\bibnamefont {Benbrik}},
  and\ \bibinfo {author} {\bibfnamefont {C.-H.}\ \bibnamefont {Chen}},\
  }\href@noop {} {\bibfield  {journal} {\bibinfo  {journal} {Phys. Rev. D}\
  }\textbf {\bibinfo {volume} {82}},\ \bibinfo {pages} {034034} (\bibinfo
  {year} {2010})}\BibitemShut {NoStop}%
\bibitem [{\citenamefont {Cao}\ \emph {et~al.}(2010{\natexlab{b}})\citenamefont
  {Cao}, \citenamefont {Heng}, \citenamefont {Wu},\ and\ \citenamefont
  {Yang}}]{PhysRevD.81.014016}%
  \BibitemOpen
  \bibfield  {author} {\bibinfo {author} {\bibfnamefont {J.}~\bibnamefont
  {Cao}}, \bibinfo {author} {\bibfnamefont {Z.}~\bibnamefont {Heng}}, \bibinfo
  {author} {\bibfnamefont {L.}~\bibnamefont {Wu}}, and\ \bibinfo {author}
  {\bibfnamefont {J.~M.}\ \bibnamefont {Yang}},\ }\href@noop {} {\bibfield
  {journal} {\bibinfo  {journal} {Phys. Rev. D}\ }\textbf {\bibinfo {volume}
  {81}},\ \bibinfo {pages} {014016} (\bibinfo {year}
  {2010}{\natexlab{b}})}\BibitemShut {NoStop}%
\bibitem [{\citenamefont {Barger}\ \emph {et~al.}(2010)\citenamefont {Barger},
  \citenamefont {Keung},\ and\ \citenamefont {Yu}}]{PhysRevD.81.113009}%
  \BibitemOpen
  \bibfield  {author} {\bibinfo {author} {\bibfnamefont {V.}~\bibnamefont
  {Barger}}, \bibinfo {author} {\bibfnamefont {W.-Y.}\ \bibnamefont {Keung}},
  and\ \bibinfo {author} {\bibfnamefont {C.-T.}\ \bibnamefont {Yu}},\
  }\href@noop {} {\bibfield  {journal} {\bibinfo  {journal} {Phys. Rev. D}\
  }\textbf {\bibinfo {volume} {81}},\ \bibinfo {pages} {113009} (\bibinfo
  {year} {2010})}\BibitemShut {NoStop}%
\bibitem [{\citenamefont {Ferrario}\ and\ \citenamefont
  {Rodrigo}(2008)}]{PhysRevD.78.094018}%
  \BibitemOpen
  \bibfield  {author} {\bibinfo {author} {\bibfnamefont {P.}~\bibnamefont
  {Ferrario}}\ and\ \bibinfo {author} {\bibfnamefont {G.}~\bibnamefont
  {Rodrigo}},\ }\href@noop {} {\bibfield  {journal} {\bibinfo  {journal} {Phys.
  Rev. D}\ }\textbf {\bibinfo {volume} {78}},\ \bibinfo {pages} {094018}
  (\bibinfo {year} {2008})}\BibitemShut {NoStop}%
\bibitem [{\citenamefont {Ferrario}\ and\ \citenamefont
  {Rodrigo}(2009)}]{PhysRevD.80.051701}%
  \BibitemOpen
  \bibfield  {author} {\bibinfo {author} {\bibfnamefont {P.}~\bibnamefont
  {Ferrario}}\ and\ \bibinfo {author} {\bibfnamefont {G.}~\bibnamefont
  {Rodrigo}},\ }\href@noop {} {\bibfield  {journal} {\bibinfo  {journal} {Phys.
  Rev. D}\ }\textbf {\bibinfo {volume} {80}},\ \bibinfo {pages} {051701}
  (\bibinfo {year} {2009})}\BibitemShut {NoStop}%
\bibitem [{\citenamefont {Bauer}\ \emph {et~al.}(2010)\citenamefont {Bauer},
  \citenamefont {Goertz}, \citenamefont {Haisch}, \citenamefont {Pfoh},\ and\
  \citenamefont {Westhoff}}]{jhep11(2010)039}%
  \BibitemOpen
  \bibfield  {author} {\bibinfo {author} {\bibfnamefont {M.}~\bibnamefont
  {Bauer}}, \bibinfo {author} {\bibfnamefont {F.}~\bibnamefont {Goertz}},
  \bibinfo {author} {\bibfnamefont {U.}~\bibnamefont {Haisch}}, \bibinfo
  {author} {\bibfnamefont {T.}~\bibnamefont {Pfoh}}, and\ \bibinfo {author}
  {\bibfnamefont {S.}~\bibnamefont {Westhoff}},\ }\href@noop {} {\bibfield
  {journal} {\bibinfo  {journal} {J. High Energy Phys.}\ }\bibinfo {volume}
  {11} (\bibinfo {year} {2010})\ \bibinfo {pages} {039}}\BibitemShut {NoStop}%
\bibitem [{\citenamefont {Cheung}\ \emph {et~al.}(2009)\citenamefont {Cheung},
  \citenamefont {Keung},\ and\ \citenamefont {Yuan}}]{Cheung2009287}%
  \BibitemOpen
  \bibfield  {author} {\bibinfo {author} {\bibfnamefont {K.}~\bibnamefont
  {Cheung}}, \bibinfo {author} {\bibfnamefont {W.-Y.}\ \bibnamefont {Keung}},
  and\ \bibinfo {author} {\bibfnamefont {T.-C.}\ \bibnamefont {Yuan}},\
  }\href@noop {} {\bibfield  {journal} {\bibinfo  {journal} {Phys. Lett. B}\
  }\textbf {\bibinfo {volume} {682}},\ \bibinfo {pages} {287 } (\bibinfo {year}
  {2009})}\BibitemShut {NoStop}%
\bibitem [{\citenamefont {Bernreuther}\ and\ \citenamefont
  {Si}(2010)}]{Bernreuther201090}%
  \BibitemOpen
  \bibfield  {author} {\bibinfo {author} {\bibfnamefont {W.}~\bibnamefont
  {Bernreuther}}\ and\ \bibinfo {author} {\bibfnamefont {Z.-G.}\ \bibnamefont
  {Si}},\ }\href@noop {} {\bibfield  {journal} {\bibinfo  {journal} {Nucl.
  Phys.}\ }\textbf {\bibinfo {volume} {B837}},\ \bibinfo {pages} {90} (\bibinfo
  {year} {2010})}\BibitemShut {NoStop}%
\bibitem [{\citenamefont {Falkowski}\ \emph {et~al.}(2013)\citenamefont
  {Falkowski}, \citenamefont {Mangano}, \citenamefont {Martin}, \citenamefont
  {Perez},\ and\ \citenamefont {Winter}}]{PhysRevD.87.034039}%
  \BibitemOpen
  \bibfield  {author} {\bibinfo {author} {\bibfnamefont {A.}~\bibnamefont
  {Falkowski}}, \bibinfo {author} {\bibfnamefont {M.~L.}\ \bibnamefont
  {Mangano}}, \bibinfo {author} {\bibfnamefont {A.}~\bibnamefont {Martin}},
  \bibinfo {author} {\bibfnamefont {G.}~\bibnamefont {Perez}}, and\ \bibinfo
  {author} {\bibfnamefont {J.}~\bibnamefont {Winter}},\ }\href@noop {}
  {\bibfield  {journal} {\bibinfo  {journal} {Phys. Rev. D}\ }\textbf {\bibinfo
  {volume} {87}},\ \bibinfo {pages} {034039} (\bibinfo {year}
  {2013})}\BibitemShut {NoStop}%
\bibitem [{\citenamefont {Aaltonen}\ \emph
  {et~al.}(2013{\natexlab{c}})\citenamefont {Aaltonen} \emph
  {et~al.}}]{Aaltonen:2013vaf}%
  \BibitemOpen
  \bibfield  {author} {\bibinfo {author} {\bibfnamefont {T.}~\bibnamefont
  {Aaltonen}}\  \emph {et~al.} (\bibinfo {collaboration} {CDF Collaboration}),\
  }\href@noop {} {\bibfield  {journal} {\bibinfo  {journal} {Phys. Rev. D}\
  }\textbf {\bibinfo {volume} {88}},\ \bibinfo {pages} {072003} (\bibinfo
  {year} {2013}{\natexlab{c}})}\BibitemShut {NoStop}%
\bibitem [{\citenamefont {Aaltonen}\ \emph
  {et~al.}(2013{\natexlab{d}})\citenamefont {Aaltonen} \emph
  {et~al.}}]{CDF11035}%
  \BibitemOpen
  \bibfield  {author} {\bibinfo {author} {\bibfnamefont {T.}~\bibnamefont
  {Aaltonen}}\  \emph {et~al.} (\bibinfo {collaboration} {CDF Collaboration}),\
  }\href@noop {} {}\bibinfo {type} {CDF Public Note}\ \bibinfo {number}
  {11035}\ (\bibinfo {year} {2013})\BibitemShut {NoStop}%
\bibitem [{\citenamefont {Abazov}\ \emph
  {et~al.}(2013{\natexlab{a}})\citenamefont {Abazov} \emph
  {et~al.}}]{Abazov:2013wxa}%
  \BibitemOpen
  \bibfield  {author} {\bibinfo {author} {\bibfnamefont {V.}~\bibnamefont
  {Abazov}}\  \emph {et~al.} (\bibinfo {collaboration} {D0 Collaboration}),\
  }\href@noop {} {\bibfield  {journal} {\bibinfo  {journal} {Phys. Rev. D}\
  }\textbf {\bibinfo {volume} {88}},\ \bibinfo {pages} {112002} (\bibinfo
  {year} {2013}{\natexlab{a}})}\BibitemShut {NoStop}%
\bibitem [{\citenamefont {Abazov}\ \emph
  {et~al.}(2013{\natexlab{b}})\citenamefont {Abazov} \emph {et~al.}}]{D0LJafb}%
  \BibitemOpen
  \bibfield  {author} {\bibinfo {author} {\bibfnamefont {V.}~\bibnamefont
  {Abazov}}\  \emph {et~al.} (\bibinfo {collaboration} {D0 Collaboration}),\
  }\href@noop {} {}\bibinfo {type} {D0 Note}\ \bibinfo {number} {6394-CONF}\
  (\bibinfo {year} {2013})\BibitemShut {NoStop}%
\bibitem [{\citenamefont {Sj{\"o}strand}\ \emph {et~al.}(2006)\citenamefont
  {Sj{\"o}strand}, \citenamefont {Mrenna},\ and\ \citenamefont
  {Skands}}]{Sjostrand:2006za}%
  \BibitemOpen
  \bibfield  {author} {\bibinfo {author} {\bibfnamefont {T.}~\bibnamefont
  {Sj{\"o}strand}}, \bibinfo {author} {\bibfnamefont {S.}~\bibnamefont
  {Mrenna}}, and\ \bibinfo {author} {\bibfnamefont {P.~Z.}\ \bibnamefont
  {Skands}},\ }\href@noop {} {\bibfield  {journal} {\bibinfo  {journal} {J.
  High Energy Phys.}\ }\bibinfo {volume} {05} (\bibinfo {year} {2006})\
  \bibinfo {pages} {026}}\BibitemShut {NoStop}%
\bibitem [{\citenamefont {Mangano}\ \emph {et~al.}(2003)\citenamefont
  {Mangano}, \citenamefont {Moretti}, \citenamefont {Piccinini}, \citenamefont
  {Pittau},\ and\ \citenamefont {Polosa}}]{Mangano:2002ea}%
  \BibitemOpen
  \bibfield  {author} {\bibinfo {author} {\bibfnamefont {M.~L.}\ \bibnamefont
  {Mangano}}, \bibinfo {author} {\bibfnamefont {M.}~\bibnamefont {Moretti}},
  \bibinfo {author} {\bibfnamefont {F.}~\bibnamefont {Piccinini}}, \bibinfo
  {author} {\bibfnamefont {R.}~\bibnamefont {Pittau}}, and\ \bibinfo {author}
  {\bibfnamefont {A.~D.}\ \bibnamefont {Polosa}},\ }\href@noop {} {\bibfield
  {journal} {\bibinfo  {journal} {J. High Energy Phys.}\ }\bibinfo {volume}
  {07} (\bibinfo {year} {2003})\ \bibinfo {pages} {001}}\BibitemShut {NoStop}%
\bibitem [{\citenamefont {Frixione}\ \emph
  {et~al.}(2007{\natexlab{a}})\citenamefont {Frixione}, \citenamefont {Nason},\
  and\ \citenamefont {Ridolfi}}]{Frixione:2007nw}%
  \BibitemOpen
  \bibfield  {author} {\bibinfo {author} {\bibfnamefont {S.}~\bibnamefont
  {Frixione}}, \bibinfo {author} {\bibfnamefont {P.}~\bibnamefont {Nason}},
  and\ \bibinfo {author} {\bibfnamefont {G.}~\bibnamefont {Ridolfi}},\
  }\href@noop {} {\bibfield  {journal} {\bibinfo  {journal} {J. High Energy
  Phys.}\ }\bibinfo {volume} {09} (\bibinfo {year} {2007}{\natexlab{a}})\
  \bibinfo {pages} {126}}\BibitemShut {NoStop}%
\bibitem [{\citenamefont {Nason}(2004)}]{Nason:2004rx}%
  \BibitemOpen
  \bibfield  {author} {\bibinfo {author} {\bibfnamefont {P.}~\bibnamefont
  {Nason}},\ }\href {\doibase 10.1088/1126-6708/2004/11/040} {\bibfield
  {journal} {\bibinfo  {journal} {J. High Energy Phys.}\ }\bibinfo {volume}
  {11} (\bibinfo {year} {2004})\ \bibinfo {pages} {040}}\BibitemShut {NoStop}%
\bibitem [{\citenamefont {Frixione}\ \emph
  {et~al.}(2007{\natexlab{b}})\citenamefont {Frixione}, \citenamefont {Nason},\
  and\ \citenamefont {Oleari}}]{Frixione:2007vw}%
  \BibitemOpen
  \bibfield  {author} {\bibinfo {author} {\bibfnamefont {S.}~\bibnamefont
  {Frixione}}, \bibinfo {author} {\bibfnamefont {P.}~\bibnamefont {Nason}},
  and\ \bibinfo {author} {\bibfnamefont {C.}~\bibnamefont {Oleari}},\ }\href
  {\doibase 10.1088/1126-6708/2007/11/070} {\bibfield  {journal} {\bibinfo
  {journal} {J. High Energy Phys.}\ }\bibinfo {volume} {11} (\bibinfo {year}
  {2007}{\natexlab{b}})\ \bibinfo {pages} {070}}\BibitemShut {NoStop}%
\bibitem [{\citenamefont {Alioli}\ \emph {et~al.}(2010)\citenamefont {Alioli},
  \citenamefont {Nason}, \citenamefont {Oleari},\ and\ \citenamefont
  {Re}}]{Alioli:2010xd}%
  \BibitemOpen
  \bibfield  {author} {\bibinfo {author} {\bibfnamefont {S.}~\bibnamefont
  {Alioli}}, \bibinfo {author} {\bibfnamefont {P.}~\bibnamefont {Nason}},
  \bibinfo {author} {\bibfnamefont {C.}~\bibnamefont {Oleari}}, and\ \bibinfo
  {author} {\bibfnamefont {E.}~\bibnamefont {Re}},\ }\href {\doibase
  10.1007/,06(2010)043} {\bibfield  {journal} {\bibinfo  {journal} {J. High
  Energy Phys.}\ }\bibinfo {volume} {06} (\bibinfo {year} {2010})\ \bibinfo
  {pages} {043}}\BibitemShut {NoStop}%
\bibitem [{\citenamefont {Antunano}\ \emph {et~al.}(2008)\citenamefont
  {Antunano}, \citenamefont {K{\"u}hn},\ and\ \citenamefont
  {Rodrigo}}]{Antunano:2007da}%
  \BibitemOpen
  \bibfield  {author} {\bibinfo {author} {\bibfnamefont {O.}~\bibnamefont
  {Antunano}}, \bibinfo {author} {\bibfnamefont {J.~H.}\ \bibnamefont
  {K{\"u}hn}}, and\ \bibinfo {author} {\bibfnamefont {G.}~\bibnamefont
  {Rodrigo}},\ }\href@noop {} {\bibfield  {journal} {\bibinfo  {journal} {Phys.
  Rev. D}\ }\textbf {\bibinfo {volume} {77}},\ \bibinfo {pages} {014003}
  (\bibinfo {year} {2008})}\BibitemShut {NoStop}%
\bibitem [{\citenamefont {Hollik}\ and\ \citenamefont
  {Pagani}(2011)}]{PhysRevD.84.093003}%
  \BibitemOpen
  \bibfield  {author} {\bibinfo {author} {\bibfnamefont {W.}~\bibnamefont
  {Hollik}}\ and\ \bibinfo {author} {\bibfnamefont {D.}~\bibnamefont
  {Pagani}},\ }\href@noop {} {\bibfield  {journal} {\bibinfo  {journal} {Phys.
  Rev. D}\ }\textbf {\bibinfo {volume} {84}},\ \bibinfo {pages} {093003}
  (\bibinfo {year} {2011})}\BibitemShut {NoStop}%
\bibitem [{\citenamefont {Manohar}\ and\ \citenamefont
  {Trott}(2012)}]{Manohar2012313}%
  \BibitemOpen
  \bibfield  {author} {\bibinfo {author} {\bibfnamefont {A.~V.}\ \bibnamefont
  {Manohar}}\ and\ \bibinfo {author} {\bibfnamefont {M.}~\bibnamefont
  {Trott}},\ }\href@noop {} {\bibfield  {journal} {\bibinfo  {journal} {Phys.
  Lett. B}\ }\textbf {\bibinfo {volume} {711}},\ \bibinfo {pages} {313 }
  (\bibinfo {year} {2012})}\BibitemShut {NoStop}%
\bibitem [{\citenamefont {K{\"u}hn}\ and\ \citenamefont
  {Rodrigo}(2012)}]{jhep012012063}%
  \BibitemOpen
  \bibfield  {author} {\bibinfo {author} {\bibfnamefont {J.}~\bibnamefont
  {K{\"u}hn}}\ and\ \bibinfo {author} {\bibfnamefont {G.}~\bibnamefont
  {Rodrigo}},\ }\href {\doibase 10.1007/JHEP01(2012)063} {\bibfield  {journal}
  {\bibinfo  {journal} {J. High Energy Phys.}\ }\bibinfo {volume} {01}
  (\bibinfo {year} {2012})\ \bibinfo {pages} {063}}\BibitemShut {NoStop}%
\bibitem [{\citenamefont {Alwall}\ \emph {et~al.}(2007)\citenamefont {Alwall},
  \citenamefont {Demin}, \citenamefont {de~Visscher}, \citenamefont {Frederix},
  \citenamefont {Herquet}, \citenamefont {Maltoni}, \citenamefont {Plehn},
  \citenamefont {Rainwater},\ and\ \citenamefont {Stelzer}}]{Alwall:2007st}%
  \BibitemOpen
  \bibfield  {author} {\bibinfo {author} {\bibfnamefont {J.}~\bibnamefont
  {Alwall}}, \bibinfo {author} {\bibfnamefont {P.}~\bibnamefont {Demin}},
  \bibinfo {author} {\bibfnamefont {S.}~\bibnamefont {de~Visscher}}, \bibinfo
  {author} {\bibfnamefont {R.}~\bibnamefont {Frederix}}, \bibinfo {author}
  {\bibfnamefont {M.}~\bibnamefont {Herquet}}, \bibinfo {author} {\bibfnamefont
  {F.}~\bibnamefont {Maltoni}}, \bibinfo {author} {\bibfnamefont
  {T.}~\bibnamefont {Plehn}}, \bibinfo {author} {\bibfnamefont {D.~L.}\
  \bibnamefont {Rainwater}}, and\ \bibinfo {author} {\bibfnamefont
  {T.}~\bibnamefont {Stelzer}},\ }\href {\doibase
  10.1088/1126-6708/2007/09/028} {\bibfield  {journal} {\bibinfo  {journal} {J.
  High Energy Phys.}\ }\bibinfo {volume} {09} (\bibinfo {year} {2007})\
  \bibinfo {pages} {028}}\BibitemShut {NoStop}%
\bibitem [{man()}]{mangano}%
  \BibitemOpen
  \href@noop {} {}\bibinfo {note} {After communication with Dr. Michelangelo
  Mangano, we realized that this shape may not have a first-principle
  analytical explanation, but rather be a combined effect from the behavior of
  the PDFs, the matrix element and the top decay kinematics. There is some
  evidence that the charge weighted rapidity distribution of the top quark is
  actually Gaussian distributed, so the second Gaussian may be just the boosts
  as part of the decay processes of the top and the $W$ boson.}\BibitemShut
  {Stop}%
\end{thebibliography}%
